\documentclass[fleqn,usenatbib]{mnras}

\usepackage{newtxtext,newtxmath}

\usepackage[T1]{fontenc}

\DeclareRobustCommand{\VAN}[3]{#2}
\let\VANthebibliography\thebibliography
\def\thebibliography{\DeclareRobustCommand{\VAN}[3]{##3}\VANthebibliography}

\usepackage{graphicx}	
\usepackage{amsmath}	
\usepackage{multicol, blindtext, multirow}
\usepackage{xcolor}
\usepackage{siunitx}
\usepackage{gensymb} 
\usepackage{caption} 
\usepackage{subcaption}      
\usepackage{float}
 \usepackage[switch]{lineno} 

\title[ The comparison of optical variability of BLS1 and NLS1 galaxies ]{  The comparison of optical variability of broad-line Seyfert 1 and narrow-line Seyfert 1 galaxies from the view of Pan-STARRS}

\author[Wang et al. 2022]{
Hongtao Wang$^{1}$\thanks{E-mail: wanghongtao@lfnu.edu.cn }, 
Chao Guo$^{1}$,
 Hongmin Cao$^{2}$,
 Yongyun Chen$^{3}$,
 Nan Ding$^{4}$, 
Xiaotong Guo$^{5}$  
\\
$^{1}$School of Science, Langfang Normal University,  Langfang Hebei 065000, China\\
$^{2}$School of Electronic and Electrical Engineering, Shangqiu Normal University, 298 Wenhua Road, Shangqiu, Henan 476000, PR China \\ 
$^{3}$ The College of Physics and Electronic Engineering, Qujing Normal University, Qujing, Yunnan 655011, China  \\
$^{4}$ School of Physical Science and Technology, Kunming University, Kunming  Yunnan 650214, PR China \\
$^{5}$ Institute of Astronomy and Astrophysics, Anqing Normal University, Anqing, Anhui 246133, PR China  \\
}

\date{Accepted XXX. Received YYY; in original form ZZZ}

\pubyear{2022}
\defcitealias{rubinur2017}{R17}
\defcitealias{hjellming1981}{HJ81}
\newcommand{\fpath}{figures/}
\begin{document}
\label{firstpage}
\pagerange{\pageref{firstpage}--\pageref{lastpage}}
\maketitle

\begin{abstract}
By means of the data sets of the  Panoramic Survey Telescope and Rapid Response System (Pan-STARRS), we investigate the relationship between the variability amplitude and  luminosity at 5100 \AA, black hole mass, Eddington ratio, $ R_{\rm Fe \, II}$ ( the ratio of the flux of Fe II line within 4435-4685   \AA ~to the broad proportion of $\rm H\beta$ line)  as well as $ R_{5007}$ (the ratio of the flux [O III] line to the total $\rm H\beta$ line) of the broad line Seyfert 1 (BLS1) and narrow line Seyfert 1 (NLS1) galaxies sample in g,r,i,z and y bands, respectively. We also analyze the similarities and differences of the variability characteristics between the BLS1 galaxies and NLS1 galaxies. The results are listed as follows.  
(1). The cumulative probability distribution of the variability amplitude shows that NLS1 galaxies are lower than that in BLS1 galaxies. (2). We analyze the dependence of the variability amplitude with the luminosity at 5100 \AA, black hole mass, Eddington ratio, $ R_{\rm Fe \,II}$ and $ R_{5007}$, respectively. We find significantly negative correlations between the variability amplitude and Eddington ratio, insignificant correlations with the luminosity at 5100 \AA. The results also show significantly positive correlations with the black hole mass and $ R_{5007}$, significantly negative correlations with $ R_{\rm Fe\, II}$ which are consistent with \citet{2017ApJ...842...96R} in low redshift bins (z<0.4) and \citet{2010ApJ...716L..31A}. (3). The relationship between the variability amplitude and the radio loudness is investigated for 155 BLS1 galaxies and 188 NLS1 galaxies. No significant  correlations are found in our results.   
\end{abstract}  




\section{Introduction}  
Active galactic nuclei (AGNs) are luminous sources in the universe with the bolometric luminosity up to $10^{48}$ $\rm erg~s^{-1}$. The   emission from AGNs spreads across the whole electromagnetic spectrum from radio to gamma-rays. Irregular variability is one of the prominent features in all  bands\citep{1997ARA&A..35..445U}. The variability timescales of AGNs are usually from a few minutes, through months, to several years \citep[e.g.,  ][]{2004IAUS..222..525I,2004ApJ...601..692V}. The variability contains abundant information, which could be utilized to trace the physical processes around the supermassive black holes (SMBHs) \citep{2017ApJ...847..132G}.      

The optical variability of AGNs has been investigated since its discovery in the 1960s . Several theoretical models are proposed to explain the variability phenomenon, such as the variation of the mass accretion rate \citep{2008MNRAS.387L..41L}, accretion disk instabilities\citep{1998ApJ...504..671K}, multiple supernovae \citep{1997MNRAS.286..271A,1998ApJ...504..671K},  micro-lensing \citep{2000A&AS..143..465H} and Poisson process \citep{2000ApJ...544..123C}.   A phenomenological model of damped random walk (DRW) has also been used to explain the variability since 1990 and is a common way of producing artificial AGN light curves for Monte Carlo simulations \citep{1987ApJS...65....1G, 2009ApJ...698..895K}.  However, the physical mechanism under AGN variability  is still not understood clearly so far.

Type-1 active galactic nuclei show a continuous range of properties of their broad-line regions (BLRs).  They have been apparent since the pioneering spectrophotometry of AGNs by Osterbrock and his collaborators \citep{1975ApJ...197L..41O, 1976ApJ...206..898O, 1977ApJ...215..733O}.  One key property is the width of the broad component of the hydrogen lines.  \citet{1984ApL....24...43G} introduced the term “narrow-line Seyfert 1” (NLS1) to refer to a prototypical AGN with narrow hydrogen lines, but whose emission-line properties were similar to the other type-1 AGNs.  The following year, \citet{1985ApJ...297..166O}  defined NLS1s as a sub-class of type-1 AGNs, with the full width at half maximum (FWHM) of $\rm H\beta$ line less than  $\rm 2000 ~ km s^{-1} $ \citep{1985ApJ...297..166O}, much stronger Fe II line than broad line Seyfert 1 (BLS1) galaxies \citep{2001A&A...372..730V}, the flux ratio of [O III] to  $\rm H\beta$ less than 3. The NLS1 galaxies usually have steeper X-ray spectra, more violent X-ray variability, lower black hole mass ($\rm10^6 -10^8 M\odot$) and higher Eddington ratio than BLS1 galaxies. Some authors show the lower black hole mass may arise from the geometrical effect of broad line region (BLR) \citep{2016IJAA....6..166L}. However, other studies suggest that the black hole masses of NLS1 galaxies are similar to those of blazars \citep{2013MNRAS.431..210C, 2016MNRAS.458L..69B}.   

The spectra of AGNs show diverse features which can be explained by several eigenvectors. It is widely accepted that the Eigenvector 1 (EV1) is the main variance of diversity. The EV1 is driven by Eddington ratio \citep{1992ApJS...80..109B, 2000ApJ...536L...5S}.  \citet{1987ApJ...321L..23W} found that the strength of the soft X-ray excess in AGNs was correlated with the strength of optical Fe II emission. \citet{1992MNRAS.256..589P} found that AGNs with strong soft X-ray excesses had narrower Balmer lines. These two correlations mean that EV1 is correlated with the soft X-ray excess.  \citet{2014Natur.513..210S} and  \citet{2015ApJ...804L..15S}  found that the Eddington ratio and orientation governed most of the diversity seen in broad-line quasar properties.
\citet{2019ApJ...882...79P}  further explained the quasar main sequence by combination of Eddington ratio, metallicity and orientation. 

Multiple studies in the 1990s reveal that NLS1 galaxies show remarkably high amplitude, rapid X-ray variability compared with BLS1 galaxies \citep{1996A&A...305...53B}. This naturally raise the question of whether the optical variability of NLS1s is greater than that of BLS1s. \citet{2004ApJ...609...69K} discovered that the answer was “no” and the optical variability was less than that of BLS1s. The discovery in \citet{2004ApJ...609...69K} was further confirmed by \citet{2010ApJ...716L..31A, 2013AJ....145...90A}  with a large sample of SDSS AGNs from Stripe 82, and by \citet{2017ApJ...842...96R} with an even larger sample from the Catalina Real-Time Transient Survey (CRTS). It is very important to note that the study of \citet{2004ApJ...609...69K} and subsequent studies confirming it all included the substantial host galaxy starlight in the photometric aperture.  The intrinsic variability amplitude of AGNs in optical band is substantially greater than what is observed through a ground-based photometric aperture \citep{2008RMxAC..32....1G}.  \citet{2009ASPC..419..388G} showed that, as expected, the ratio of the luminosity of an AGN relative to the luminosity of host galaxy increased strongly with the Eddington ratio. This means that BLS1 galaxies have more host galaxy starlight contamination than NLS1 galxies. Thus, as  \citet{2009ASPC..419..388G}  pointed out, the real differences of variability amplitude in optical band between the BLS1s and NLS1 galaxies were actually much greater than it seems. \citet{2018ApJ...866...74S}  explored the evolution of variability in g band on timescales of weeks to years by Stripe 82 quasars along the quasar main sequence and found the variability was sensitive to $\rm L_{bol}$ which might be a new "Standard Candle" to probe cosmology.  \citet{2021ApJ...911..148K} found that more variable quasars/type 1 AGNs in UV/optical band had stronger emission lines (e.g., CIV, Mg II and [OIII]5007 emission lines) using quasars in SDSS Stripe 82, which indicated a causal link between disk turbulence and emission line production. 

In order to understand the peculiar characteristics of NLS1 galaxies, some authors investigate the similarities/differences between NLS1 galaxies and BLS1 galaxies from the view of optical variability.  \citet{2010ApJ...716L..31A}  found the variability amplitude was correlated with the emission line parameters ( e.g., Fe II emission line, $\rm[OIII]$ line and $\rm H\beta$ line), and anti-correlated with Eddington ratio based on the multi-epoched photometric data in Stripe 82 region of the Sloan Digital Sky Survey (SDSS).  \citet{2013AJ....145...90A} showed the ensemble structure function of NLS1 galaxies was similar to that of the BLS1 galaxies on the timescales larger than 10 days, and smaller variability amplitude compared with BLS1 galaxies. By means of the optical data sets in V band from the CRTS spanning 5–9 years, \citet{2017ApJ...842...96R}  found the variability amplitude was anti-correlated with the continuous spectrum luminosity and Eddington ratio, but positively correlated with the width of $\rm H\beta$ line.  

The Pan-STARRS, starting its observations since 2014, is a system for wide-field astronomical imaging operated by the Astronomy Institute at Hawaii University. The current data release was updated in January 2019. The time baseline of data sets is over five years. Compared with PTF/iPTF/ZTF and CRTS, Pan-STARRS has higher cadence, much deeper limiting magnitude in five bands (g,r,i,z and y ) and smaller photometric errors. The limiting magnitude is similar with SDSS which is only focused on about Stripe 82 region.   
\citet{LiuHY2019ApJS} constructed a homogeneous and complete sample of 14 584 BLS1 galaxies at z < 0.35 based on the SDSS DR7 data release,  which is the largest one so far. \citet{Rakshit2017} constructed a 11 101 NLS1 galaxies sample with z < 0.8 from the SDSS Data Release 12 (DR12; Alam et al. 2015), which is five times larger than previous NLS1 samples. Based on the two samples and the available Pan-STARRS data sets in g,r,i,z and y bands,  we further carry out a comparative analysis between the variability and many physical parameters of the NLS1 galaxies and BLS1 galaxies for exploring the peculiar characteristics of NLS1 galaxies. 
 
This paper is arranged as follows. In Section \ref{Data}, we describe the sample of NLS1 galaxies and BLS1 galaxies and the photometric reduction of the data sets. The method in this work is presented in Section \ref{sec:method}. The results and discussions are listed in Section \ref{sec:results}.    Finally, the  conclusions are drawn in Section \ref{sec:summary}.

\section{ Sample and Data } \label{Data}
\subsection{ NLS1 and BLS1 sample } 
  In order to investigate the similarities and differences of the variability characteristics in the largest sample of 11 101 NLS1 galaxies at z< 0.8 and 14 584 BLS1 galaxies  at z < 0.35 \citep{Rakshit2017, LiuHY2019ApJS},  we cross match the two samples in redshift - luminosity plane (luminosity bin = 0.01, redshift bin=0.002),  3192 BLS1 galaxies and 3194 NLS1 galaxies are found. The two dimension KS test \citep{Peacock1983, Fasano1987}  shows the KS D statistics value of 0.009 and the p value larger than 0.999.    

\begin{figure}
    \centering
    \includegraphics[width=1\columnwidth]{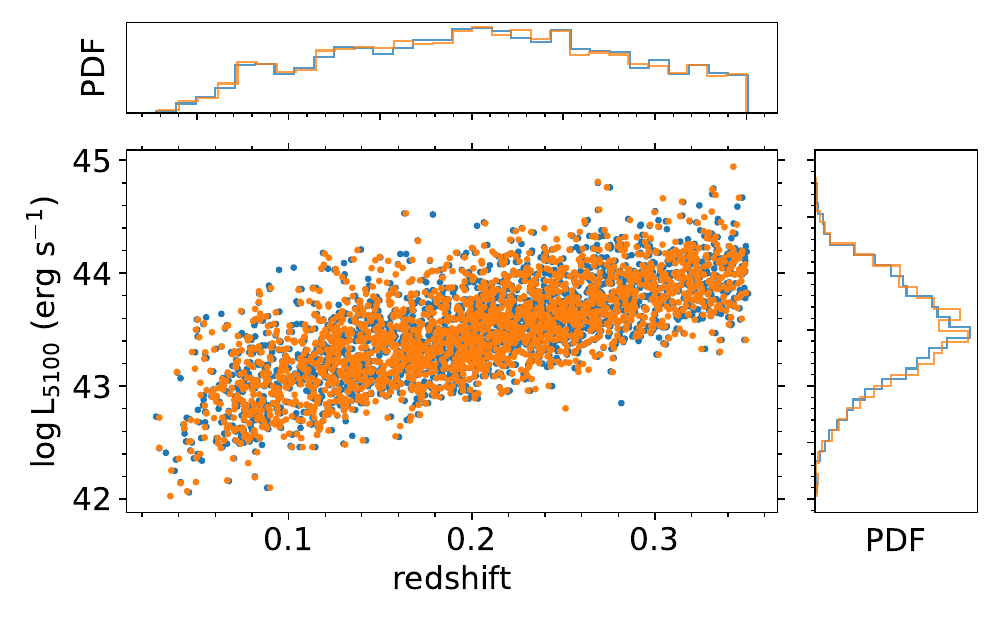}
    \caption{The luminosity at 5100 \AA ~and redshift distribution for 2530 NLS1 and 2472 BLS1 galaxies in g band. The marginal distribution of redshift (top) and luminosity (right) is  also shown.  Both BLS1 galaxies (yellow step) and NLS1 galaxies(blue step) have similar distributions.  }
    \label{fig:alpha}
\end{figure}

\subsection{ Photometric  data reduction }  
 Pan-STARRS1 (PS1) is the first part of Pan-STARRS to be completed and is the basis for both Data Releases 1 and 2 (DR1 and DR2). The PS1 survey  images the sky in five broadband filters (g, r, i, z and y bands) by a 1.8-meters telescope. First, we cross match our BLS1 and NLS1 galaxies sample with the Pan-SRARRS DR1 and DR2, then obtain the NLS1 sample of light curve in g, r, i, z and y bands. The sigma-clipping algorithm is utilized to remove the outliers which show large photometric uncertainty and cosmic-ray in the data sets. The outliers are defined as the values deviated from the five times standard deviation. The procedure is iterated until all the outliers are eliminated. We only choose the light curve more than five data sets and the time baseline longer than two years. Finally, we obtain 2530 NLS1 galaxies and 2472 galaxies in g band,  2841 NLS1 galaxies and 2798 BLS1 galaxies in r band,  2798 NLS1 galaxies and 2796 BLS1 galaxies in i band,  2942 NLS1 galaxies and 2918 BLS1 galaxies in z band and 2966 NLS1 galaxies and 2960 BLS1 galaxies in y band, respectively. The result in g band is shown in Figure 1. The two dimension KS test show the p-values are all larger than 0.99 and D value almost 0.01 in aforementioned sample. 

\section{ Method } \label{sec:method}  
Through the variance analysis of observed data sets, we calculate the variability amplitude which excludes the measurement errors in the light curves \citep{2007AJ....134.2236S,2017ApJ...842...96R,2019MNRAS.483.2362R}.      
The expression is as follows,  
\begin{equation} 
\Sigma=\sqrt{\frac{1}{N-1}\sum_{i=1}^{N}(m_i - <m>)^2},   
\end{equation}  
\noindent  
where $m_i$ is the i-th magnitude and $<m>$ is the average magnitude in the data sets. The error  $\epsilon$ is expressed as   
\begin{equation}     
\epsilon^2=\frac{1}{N}\sum_{i=i}^{N}{\epsilon_{i}^{2} }        
\end{equation}
\noindent
in which $\epsilon_i$ is the i-th error. Finally, the expression of the variability amplitude  is 
\begin{equation}    
\sigma_m  = 
  \begin{cases}
    \sqrt{\Sigma^2 - \epsilon^2},  & \quad \text{if } \Sigma>\epsilon,\\
     0,                            & \quad  \text{otherwise.}\\
  \end{cases}
\end{equation}   

 Compared with other methods, it calculates the variability amplitude directly,  is not dependent on models, and also takes the impact of photometric errors into account.  

The Seyfert galaxies are usually less luminous than quasars and detected their host galaxies, hence the starlight contamination from the host galaxies of NLS1 galaxies is not negligible.  We revise it by the method of \citet{2011ApJS..194...45S}  which provided us an empirical fitting formula of the average host contamination. The expression is 
\begin{equation}   
\dfrac{L_{5100,host}}{L_{5100,QSO}} = 0.8052- 1.5502x+0.9121x^2-0.1577x^3  , 
\end{equation}  
\noindent
in which  $x+44 \equiv log(L_{5100,total} / erg s^{-1}) < 45.053$. We extend it to $x < 0$ due to the low luminosity of NLS1 and BLS1 galaxies.  

\section{ Results } \label{sec:results} 

\subsection{ The cumulative probability distribution of the variability amplitude $\sigma_m$ of BLS1 and NLS1 galaxies in g, r, i, z and y bands,respectively.  } \label{Spectral signatures}      
In order to compare the difference of the variability amplitude, we investigate the cumulative probability distribution of the variability amplitude $\sigma_m$  in the BLS1 and NLS1 galaxies sample. The results are shown in Figure 2. We find that the variability amplitude in BLS1 galaxies sample is larger than that in NLS1 galaxies sample in g, r, i, z and y bands, respectively. The median value of the variability amplitude in BLS1 and NLS1 galaxies sample is 0.142 mag and 0.119 mag in g band,  0.137 mag and 0.114 mag in r band, 0.130 mag and 0.112 mag in i band,  0.125 mag and 0.103 mag in z band,  0.132 mag and 0.115 mag in y band, respectively. 
   
\citet{2010ApJ...716L..31A,2013AJ....145...90A} found the variability amplitude of 58 NLS1 galaxies was systematically smaller than that of 217 BLS1 galaxies by the multi-epoch photometric data in the Stripe 82 region of the SDSS. \citet{2017ApJ...842...96R} further found the variability amplitude of NLS1 galaxies is smaller than BLS1 galaxies in V band from the CRTS spanning 5-9 years. Our results are consistent with the situations mentioned above and further verify the characteristics of weaker variability in NLS1 galaxies.       
\begin{figure*}
\begin{subfigure}{.33\textwidth}
  \centering
  \includegraphics[width=.98\linewidth]{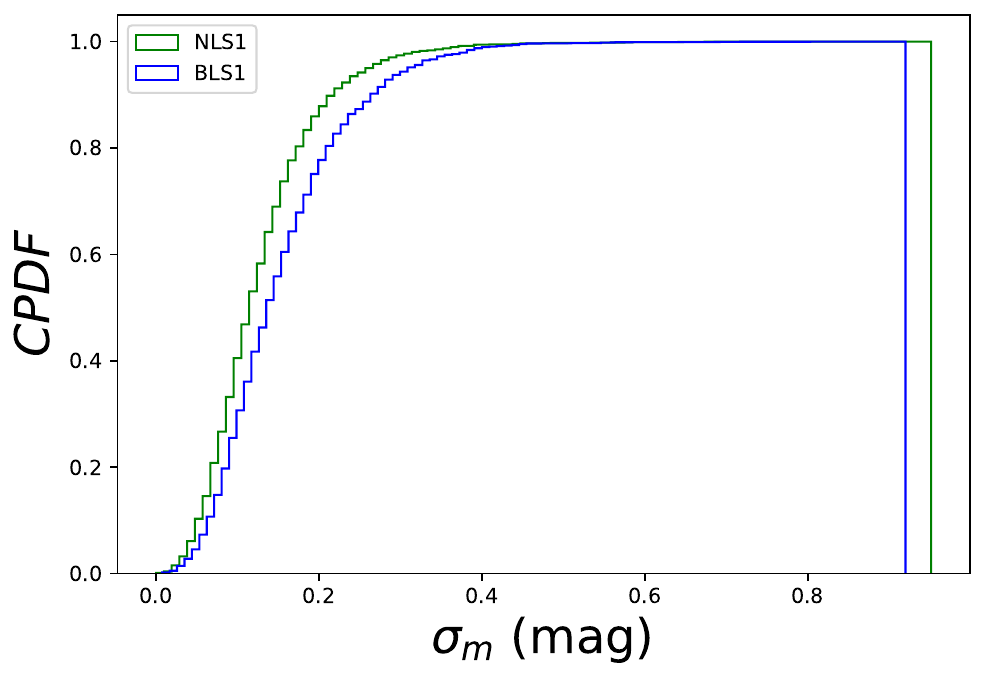}  
  \caption{ The results in g band }         
  \label{fig:sub-first}
\end{subfigure}       
\hfill
\begin{subfigure}{ 0.33\textwidth}  
  \centering
  \includegraphics[  width=.95\linewidth]{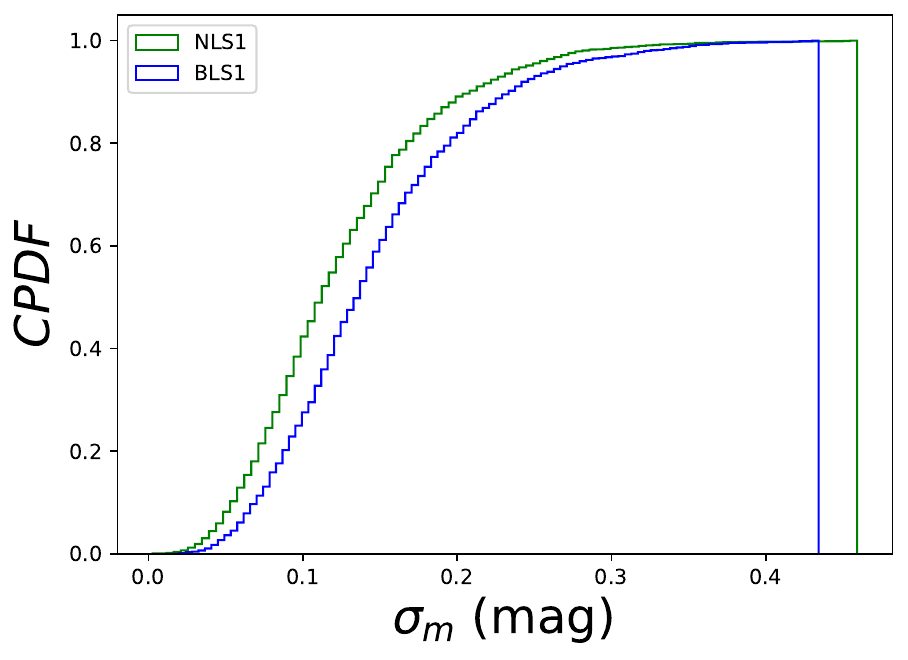}  
  \caption{ The results in r band }
  \label{fig:sub-second}
\end{subfigure}
\hfill
\begin{subfigure}{.33\textwidth}  
  \centering
  \includegraphics[width=.98\linewidth]{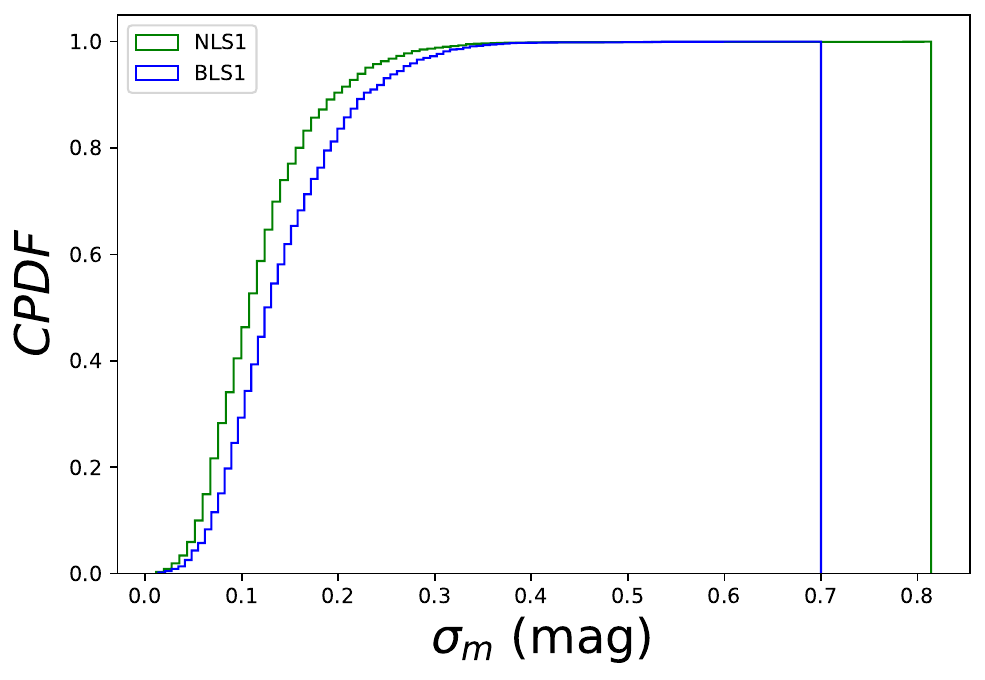}  
  \caption{ The results in i band }
  \label{fig:sub-second}
\end{subfigure}
\hfill
\begin{subfigure}{.33\textwidth}
  \centering
  \includegraphics[width=.98\linewidth]{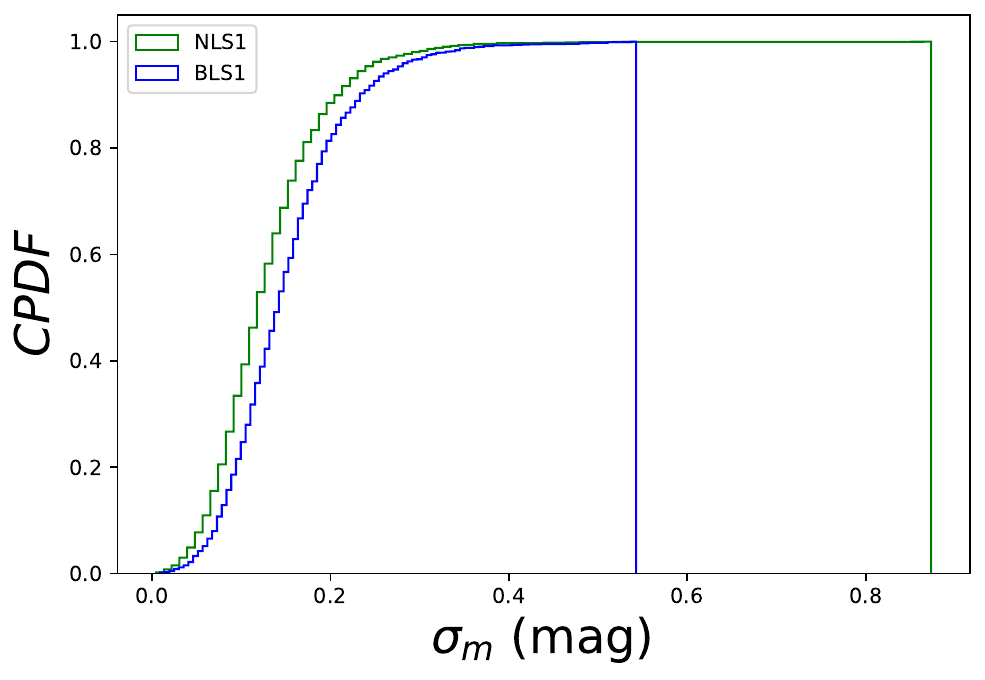}  
  \caption{  The results in z band}
  \label{fig:sub-second}
\end{subfigure}
\begin{subfigure}{.33\textwidth}
  \centering
  \includegraphics[width=.98\linewidth]{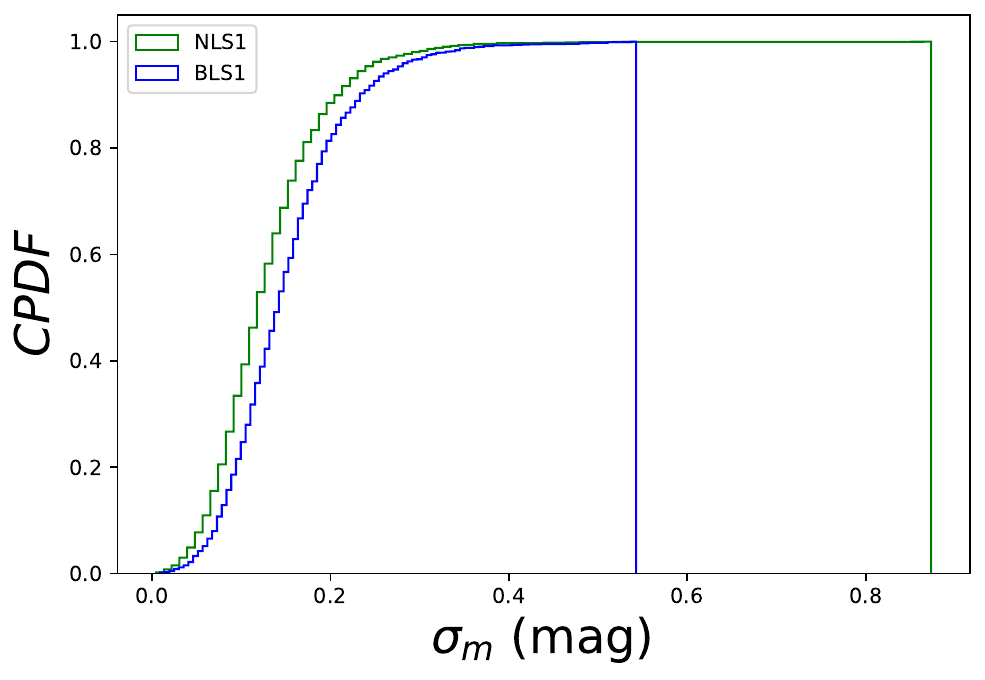}  
  \caption{ The results in y band  }
  \label{fig:sub-second}   
\end{subfigure}
\caption{  The cumulative probability distribution function of variability amplitude of BLS1 and NLS1 galaxies in g, r, i, z and y bands, respectively.  }
\label{fig:fig}
\end{figure*}

\begin{figure*}     
\begin{subfigure}{.33\textwidth}    
  \centering
  \includegraphics[width=.98\linewidth]{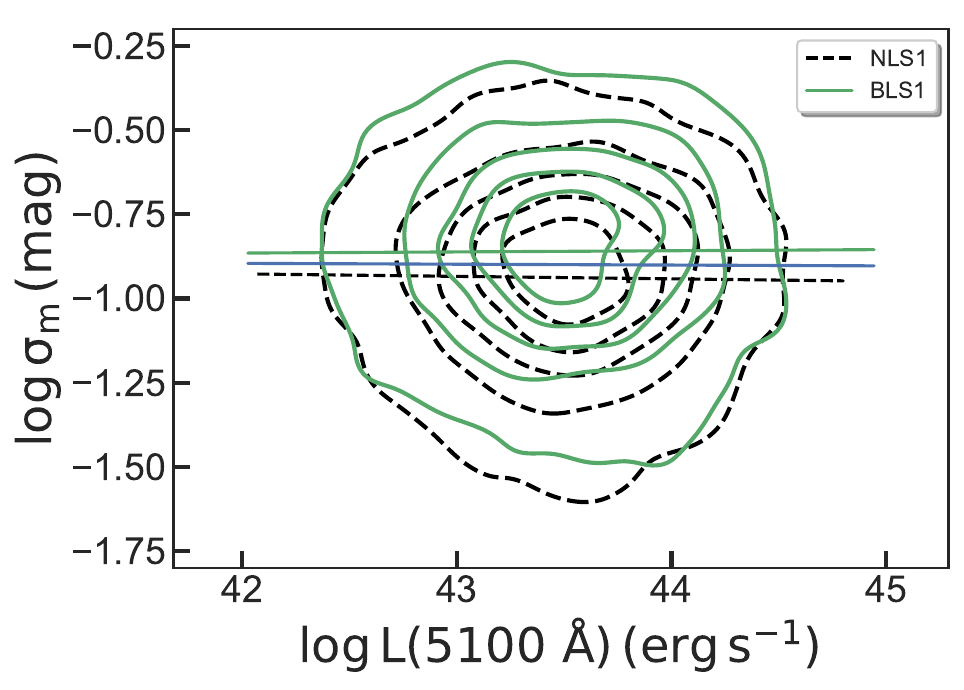}  
  \caption{   }
  \label{fig:sub-first}
\end{subfigure}
\begin{subfigure}{.33\textwidth}
  \centering
  \includegraphics[width=.98\linewidth]{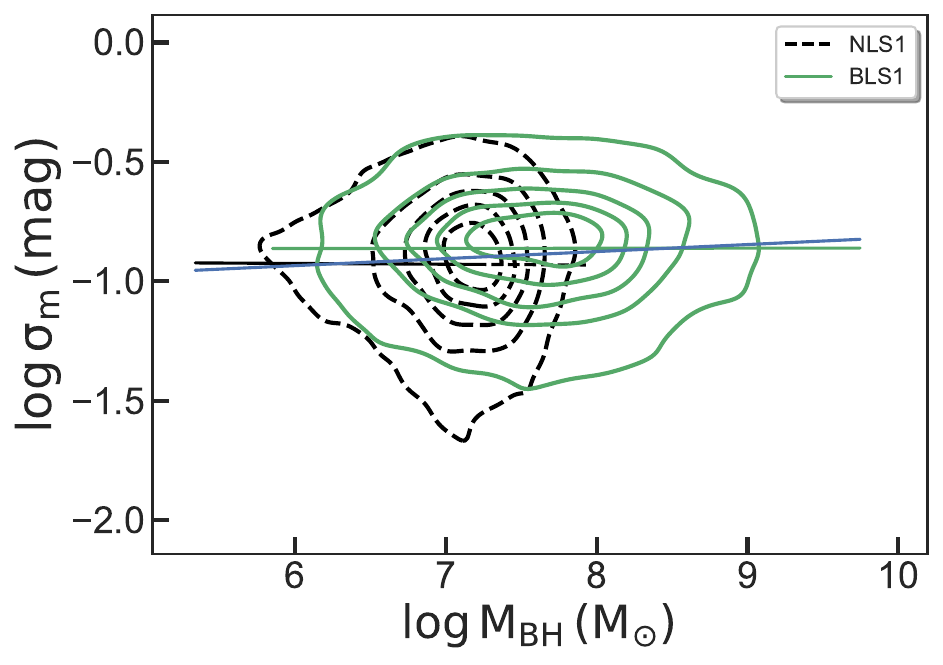}  
  \caption{   }
  \label{fig:sub-second}
\end{subfigure}
\hfill
\begin{subfigure}{.33\textwidth}
  \centering
  \includegraphics[width=.98\linewidth]{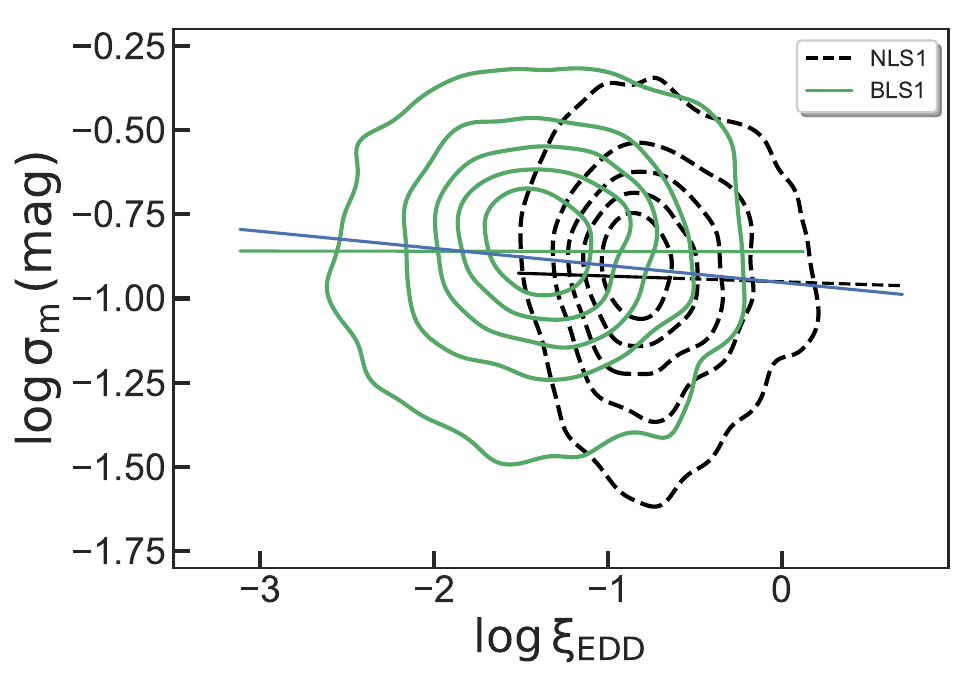}  
  \caption{   }
  \label{fig:sub-second}
\end{subfigure}
\hfill
\begin{subfigure}{.33\textwidth}
  \centering
  \includegraphics[width=.98\linewidth]{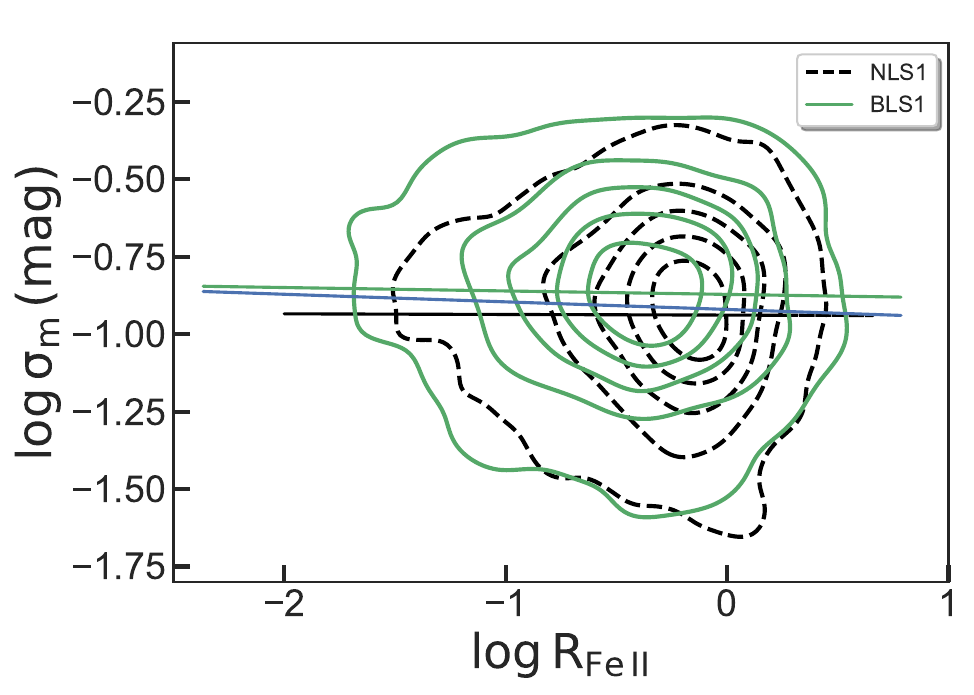}  
  \caption{   }
  \label{fig:sub-second}
\end{subfigure}
\begin{subfigure}{.33\textwidth}
  \centering
  \includegraphics[width=.98\linewidth]{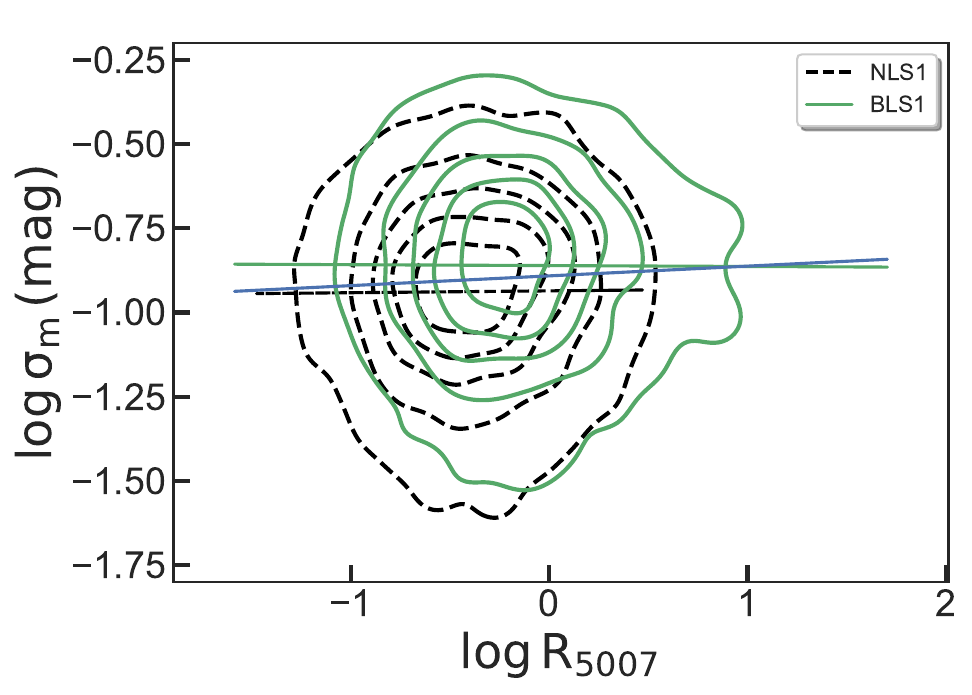}  
  \caption{  }
  \label{fig:sub-second}   
\end{subfigure}   
\caption{  The relationship between variability amplitude and the luminosity at 5100 \AA, black hole mass, Eddington ratio, $ R_{\rm Fe \,II}$ and $R_{5007}$ of BLS1 and NLS1 galaxies in g band.    }
\label{fig:fig}
\end{figure*}

\begin{figure*} 
\begin{subfigure}{.33\textwidth}
  \centering
  \includegraphics[width=.98\linewidth]{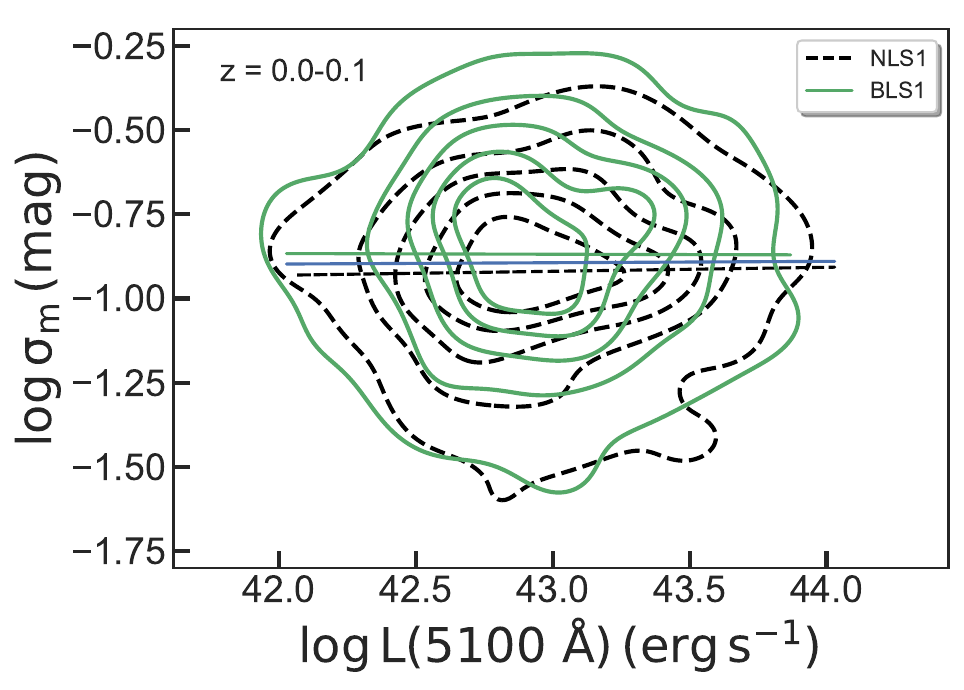}  
  \caption{     }
  \label{fig:sub-first}
\end{subfigure}
\begin{subfigure}{.33\textwidth}
  \centering
  \includegraphics[width=.98\linewidth]{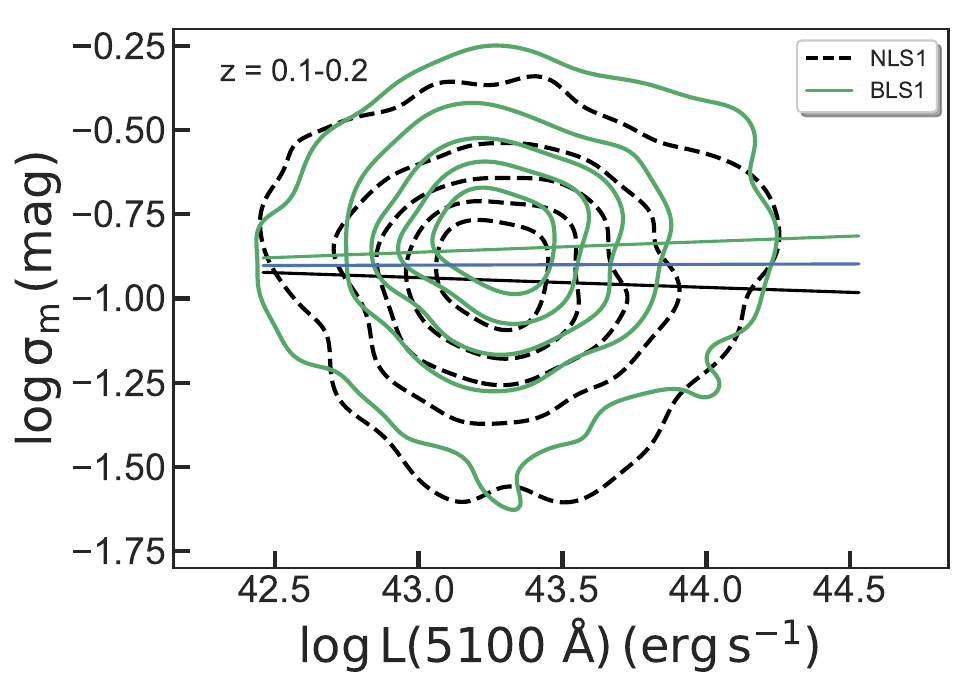}  
  \caption{     }
  \label{fig:sub-second}
\end{subfigure}
\hfill
\begin{subfigure}{.33\textwidth}
  \centering
  \includegraphics[width=.98\linewidth]{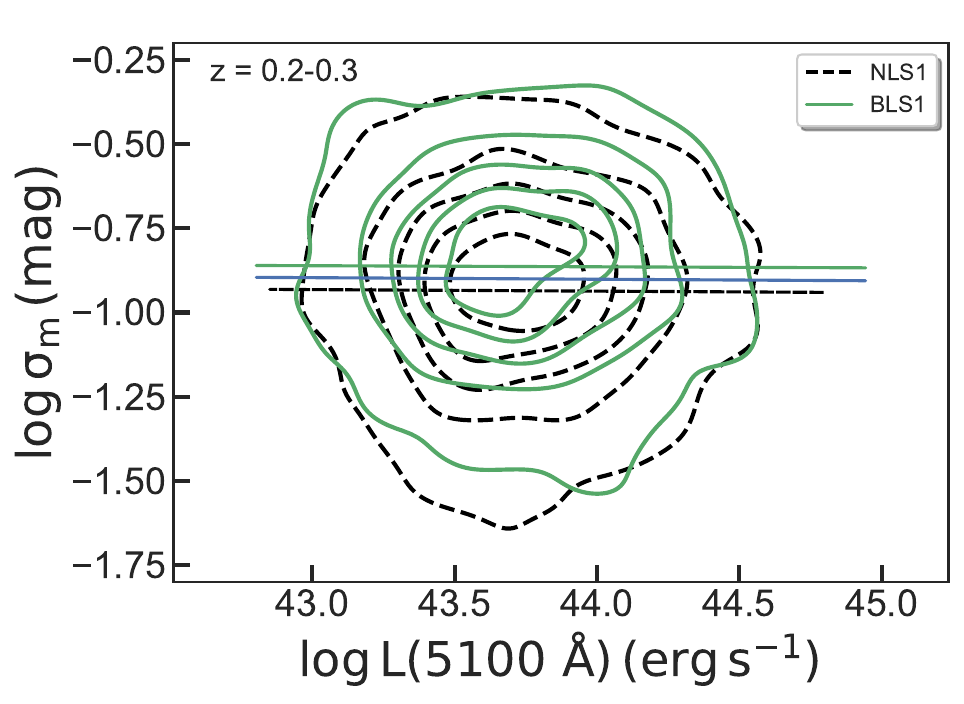}  
  \caption{    }
  \label{fig:sub-second}
\end{subfigure}
\hfill
\begin{subfigure}{.33\textwidth}
  \centering
  \includegraphics[width=.98\linewidth]{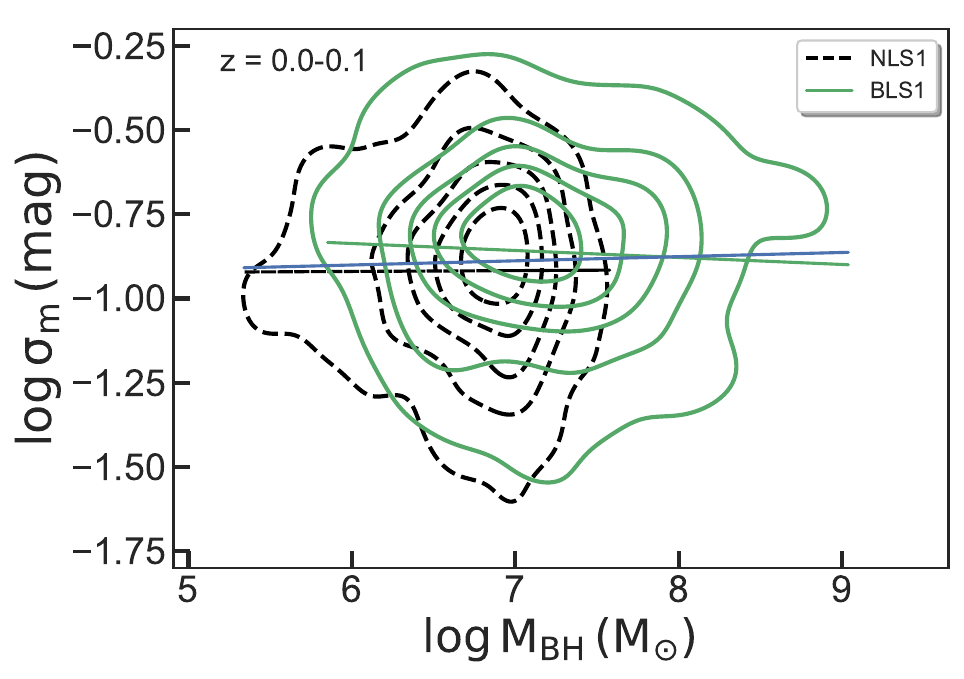}  
  \caption{     }
  \label{fig:sub-second}
\end{subfigure}
\begin{subfigure}{.33\textwidth}
  \centering
  \includegraphics[width=.98\linewidth]{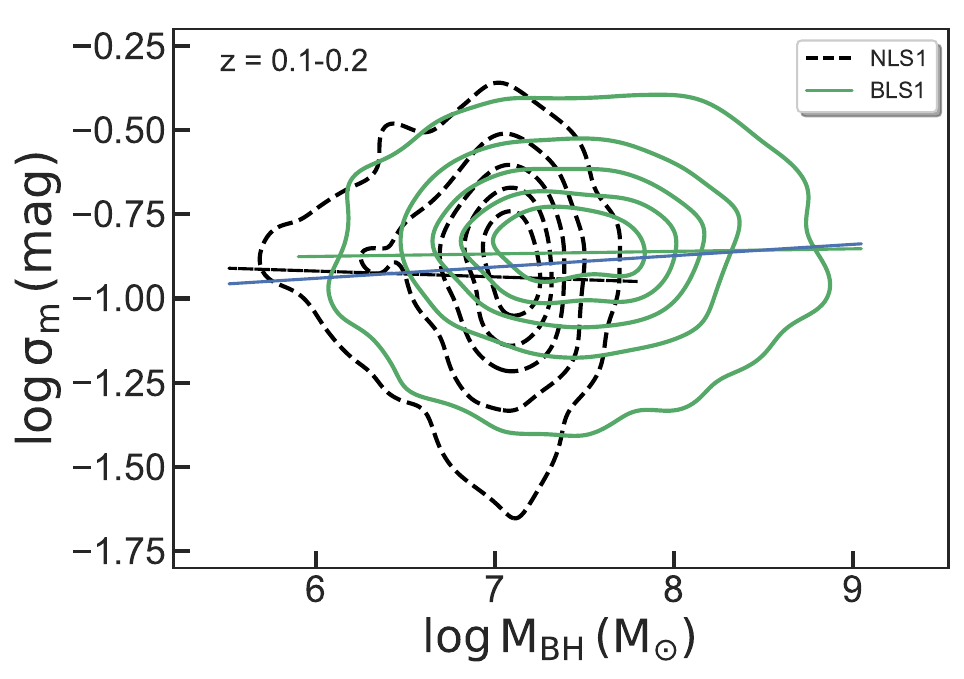}  
  \caption{    }
  \label{fig:sub-second}   
\end{subfigure} 
\begin{subfigure}{.33\textwidth}
  \centering
  \includegraphics[width=.98\linewidth]{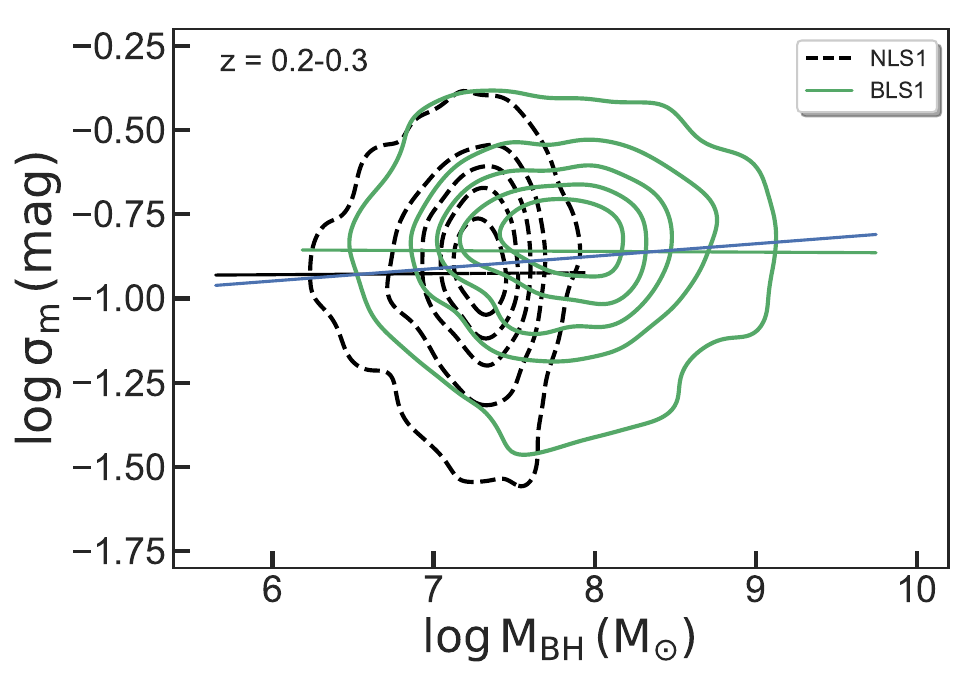}  
  \caption{    }
  \label{fig:sub-second}   
\end{subfigure} 
\begin{subfigure}{.33\textwidth}
  \centering
  \includegraphics[width=.98\linewidth]{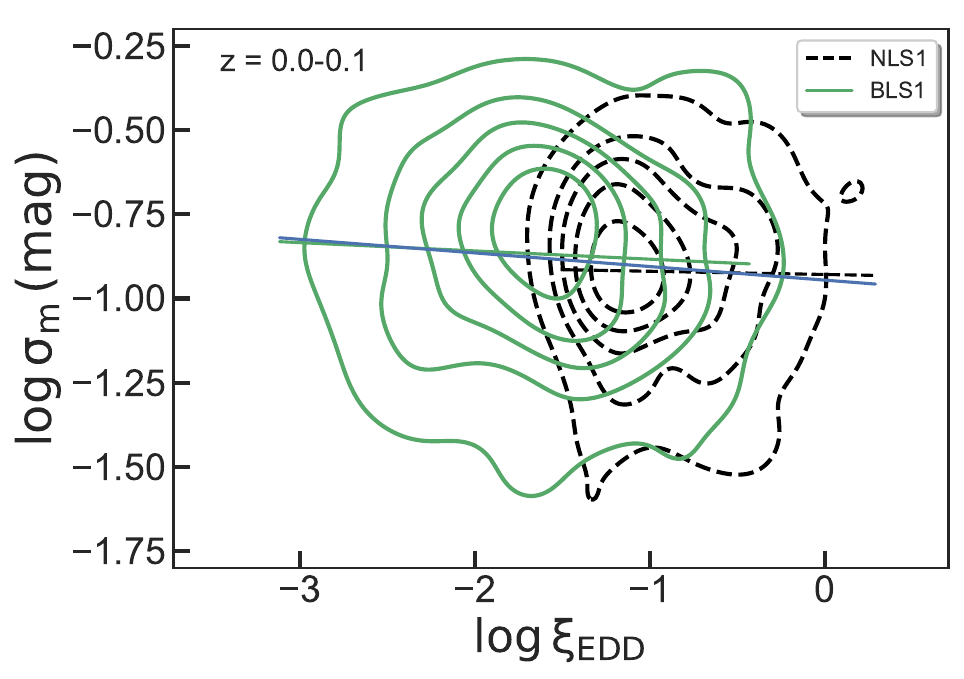}  
  \caption{     }
  \label{fig:sub-second}
\end{subfigure}
\begin{subfigure}{.33\textwidth}
  \centering
  \includegraphics[width=.98\linewidth]{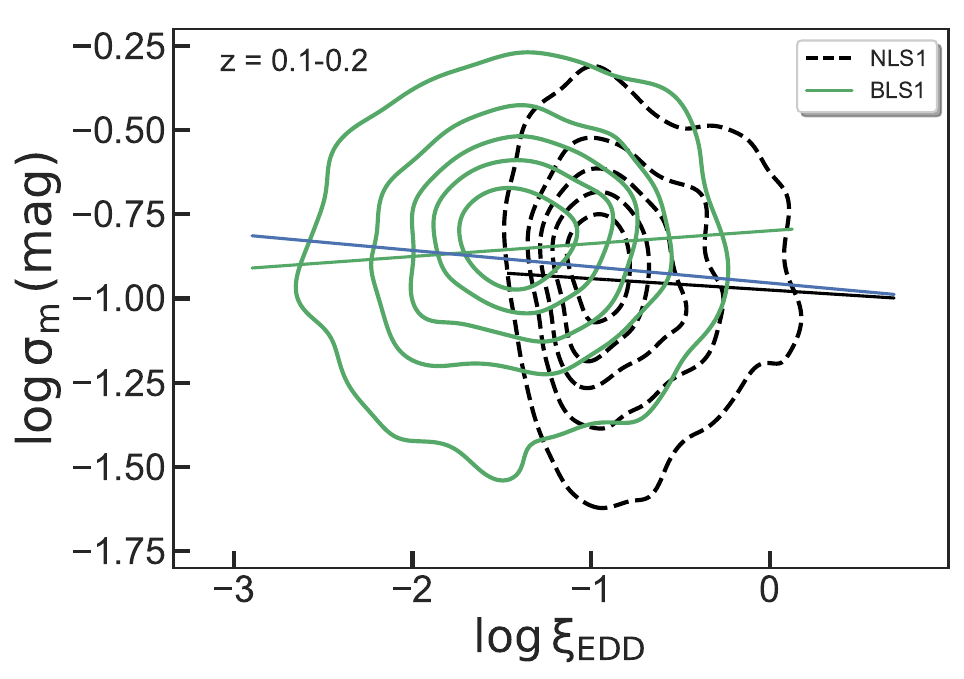}  
  \caption{    }
  \label{fig:sub-second}   
\end{subfigure} 
\begin{subfigure}{.33\textwidth}
  \centering
  \includegraphics[width=.98\linewidth]{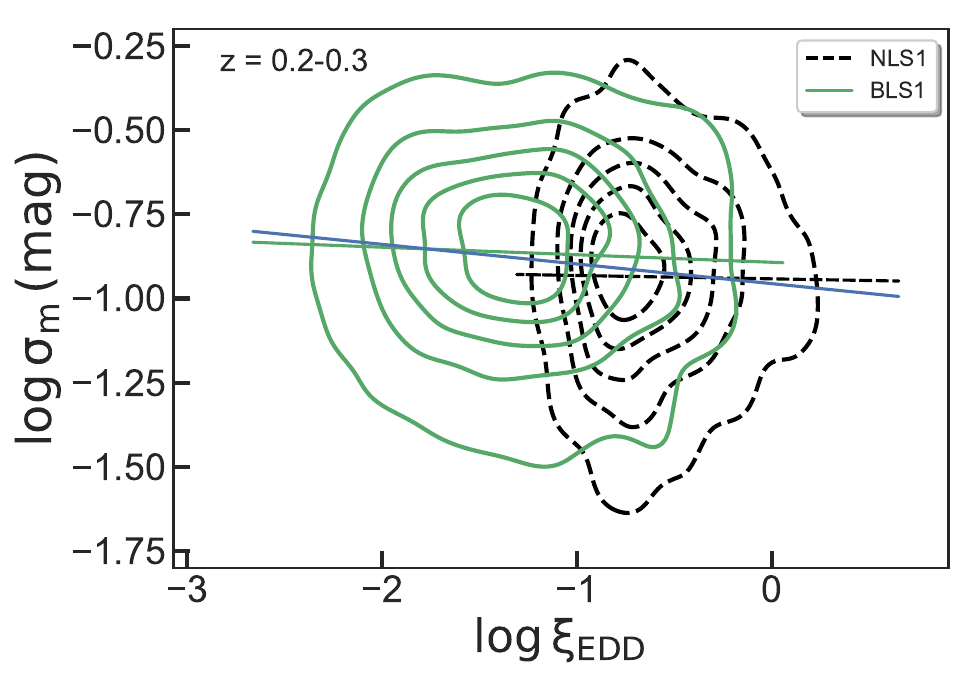}  
  \caption{    }
  \label{fig:sub-second}   
\end{subfigure} 
\begin{subfigure}{.33\textwidth}
  \centering
  \includegraphics[width=.98\linewidth]{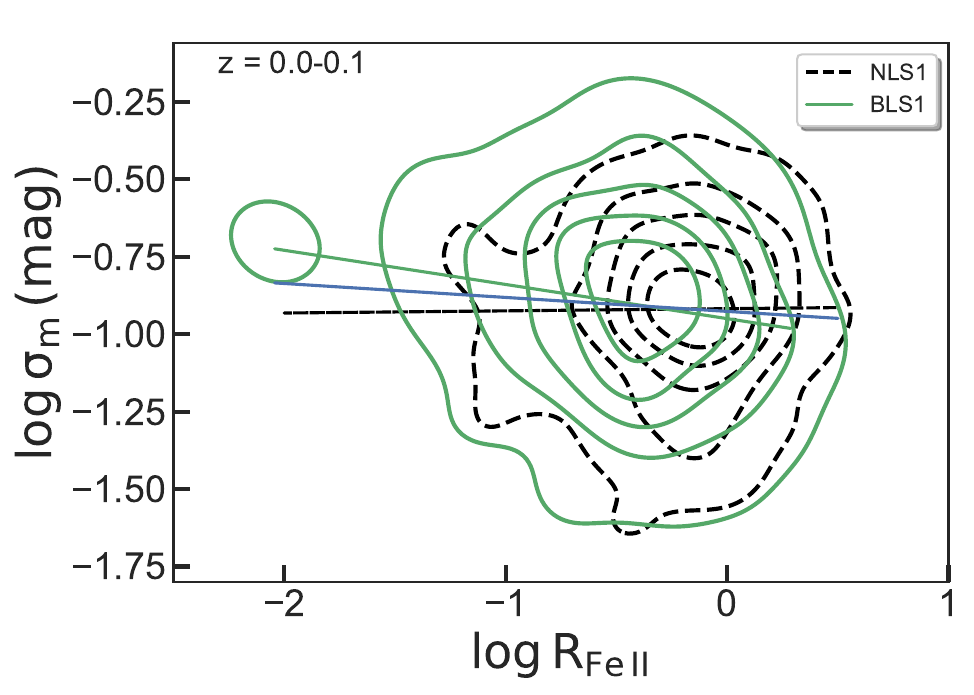}  
  \caption{     }
  \label{fig:sub-second}
\end{subfigure}
\begin{subfigure}{.33\textwidth}
  \centering
  \includegraphics[width=.98\linewidth]{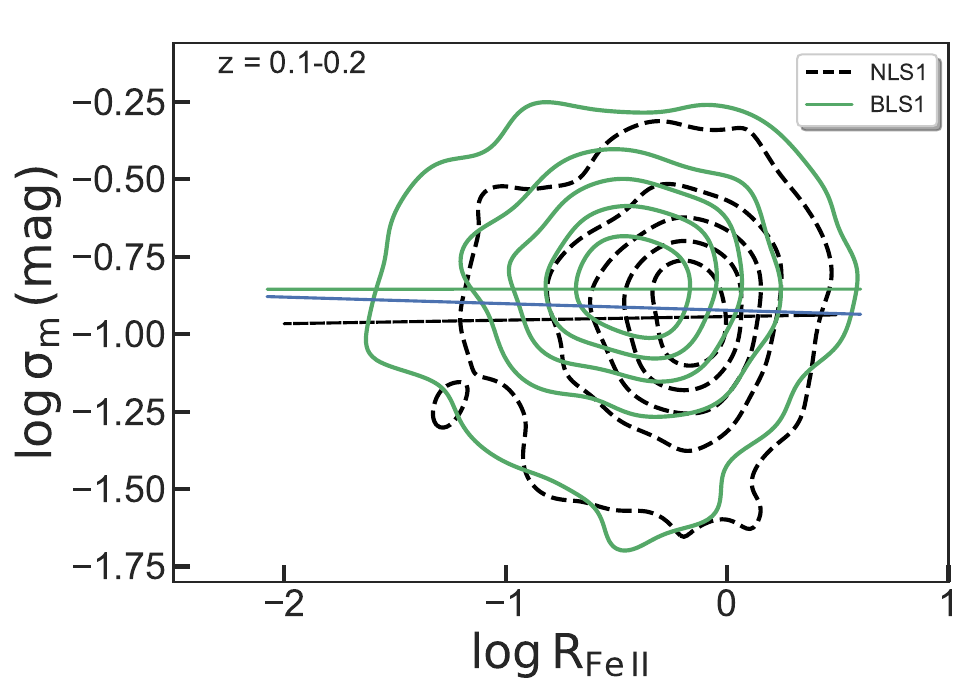}  
  \caption{   }
  \label{fig:sub-second}   
\end{subfigure} 
\begin{subfigure}{.33\textwidth}
  \centering
  \includegraphics[width=.98\linewidth]{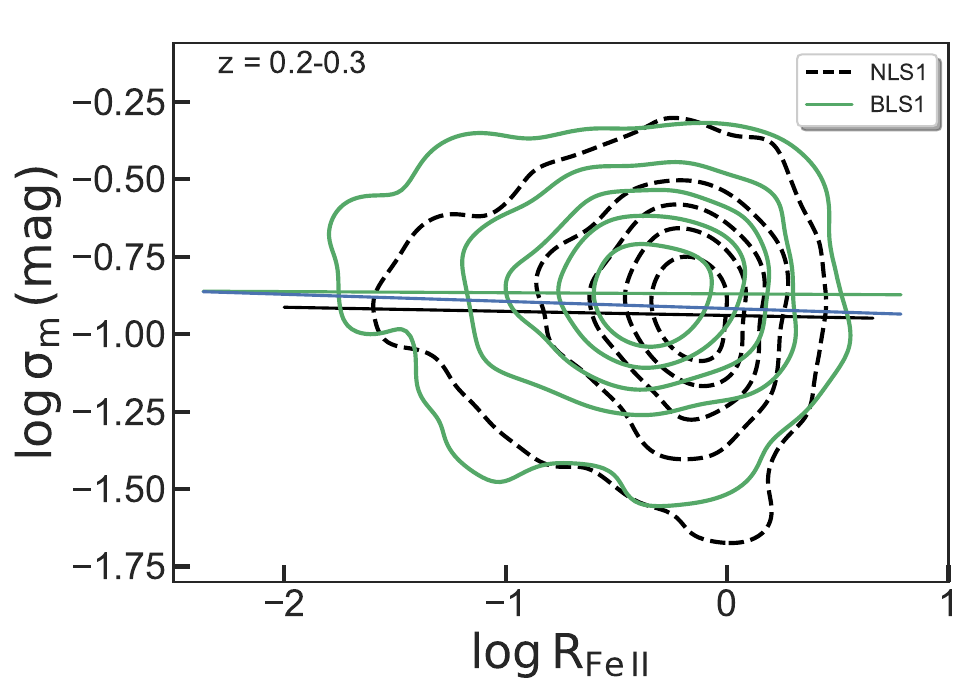}  
  \caption{   }
  \label{fig:sub-second}   
\end{subfigure}  
\begin{subfigure}{.33\textwidth}
  \centering
  \includegraphics[width=.98\linewidth]{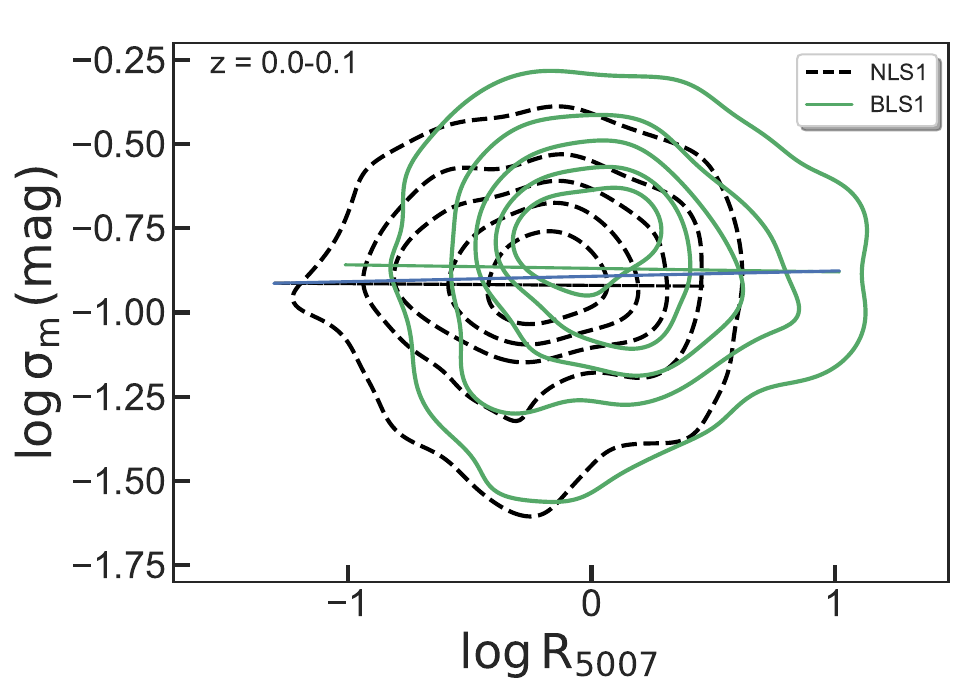}  
  \caption{   }
  \label{fig:sub-second}
\end{subfigure}
\begin{subfigure}{.33\textwidth}
  \centering
  \includegraphics[width=.98\linewidth]{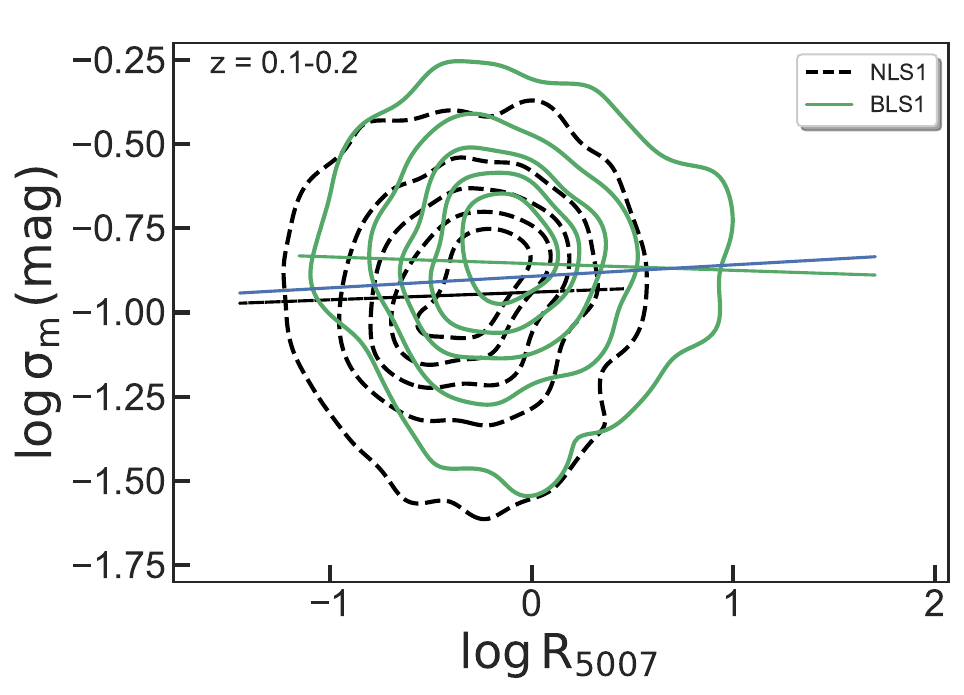}  
  \caption{  }
  \label{fig:sub-second}   
\end{subfigure} 
\begin{subfigure}{.33\textwidth}
  \centering
  \includegraphics[width=.98\linewidth]{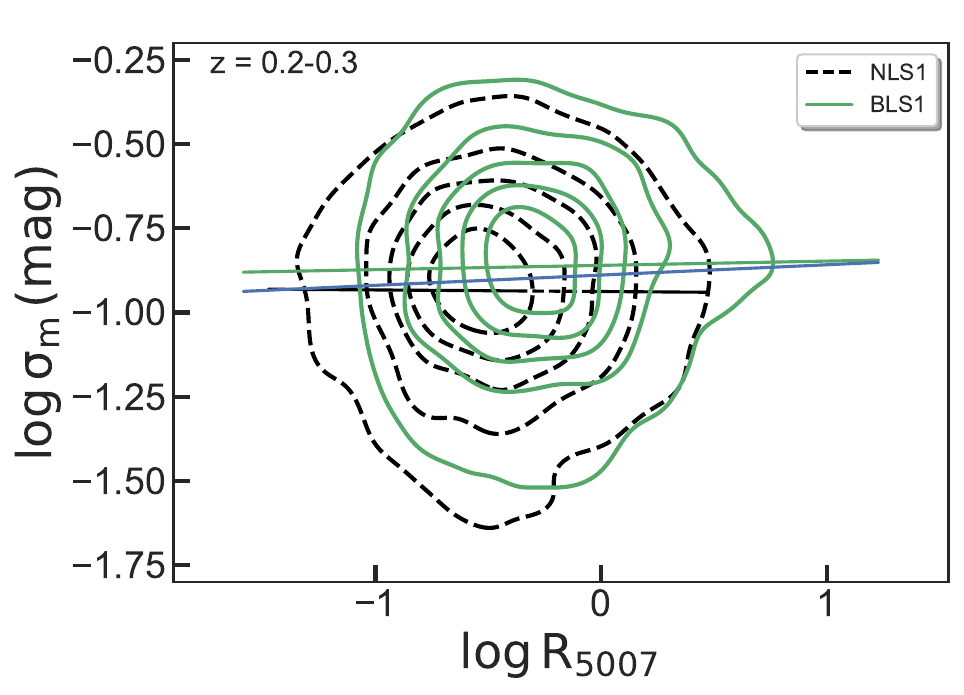}  
  \caption{   }
  \label{fig:sub-second}   
\end{subfigure} 
\caption{ The variability in g band dependence with luminosity at 5100 \AA, black hole mass, Eddington ratio, $\rm R_{Fe \,II}$ and $R_{5007}$ in the  redshift bin 0-0.1, 0.1-0.2 and 0.2-0.35 for BLS1 and NLS1 galaxies.   }
\label{fig:fig}
\end{figure*}

\subsection{ The relationship between the variability amplitude and the physical parameters   } 
In this section, we analyze the dependence of the variability amplitude with the luminosity at 5100 \AA,  black hole mass, Eddington ratio, $\rm R_{Fe \,II}$ and $R_{5007}$, respectively. The results in g band as an example are shown in Figure 3. 
The fitting results by the least square method, Spearman correlation coefficient and p-value are listed in Table 1, respectively.     

\subsubsection{ The dependence of variability with luminosity at 5100 \AA }     
We investigate the relationship between the variability amplitude and the luminosity at 5100 \AA. The result in g band is shown in the upper left panel (a) of Figure 3. The dashed black  contour and solid blue contour present the NLS1 galaxies and BLS1 galaxies sample, respectively. No noticeable differences were found between them. We further analyze the correlation by the least square method. The dashed black fitting line,  solid green fitting line and solid blue line present the results of NLS1 sample, BLS1 sample and all sample, respectively.  No significant correlations are found in our results.       

 \citet{2010ApJ...716L..31A}  had also shown an insignificant correlation between the variability amplitude and luminosity.  \citet{2017ApJ...842...96R} found significantly positive correlation at lower redshift (z<0.4) between the variability amplitude and luminosity at 5100 \AA, but significant anti-correlation at higher redshift (0.4 < z <0.8).  \citet{2022RAA....22a5014W} analyzed the relationship of the 11101 NLS1 galaxies from Pan-STARRS which showed mild anti-correlation. Compared with aforementioned results, the insignificant correlation between variability amplitude and luminosity in lower redshift are needed to further investigate again in the future.  

\subsubsection{ The dependence of variability with black hole mass   } 
The black hole mass of NLS1 and BLS1 galaxies in our work is from the catalog of \citet{Rakshit2017} and \citet{LiuHY2019ApJS}, respectively. They estimated them by the virial relation using the virial motion of the clouds in broad line region around the central black hole. Then, the 
$\rm M_{BH}$ can be expressed as  
\begin{equation}        
    \label{eq:kings}    
    \centering      
    M_{BH} =  fR_{BLR}\Delta v^{2}/G   
\end{equation}     
in which $f$ is the scale factor which is related to the structure and dynamics of the BLR, $f =0.75$ in this work. $\delta v$ is the FWHM of the $\rm H\beta$ emission line. G is the gravitational constant. The relationship between the variability amplitude and black hole mass is shown in the upper middle panel of Figure 3. The dashed black contour and solid blue contour present the NLS1 galaxies and BLS1 galaxies sample, respectively. The $\rm M_{BH}$ of NLS1 galaxies is obviously smaller than that in BLS1 galaxies which is consistent with the results in \citet{Decarli2008} and \citet{2017ApJ...842...96R}. The dashed black fitting line, solid green fitting line and solid blue line present the results of the NLS1 sample, BLS1 sample and all sample by least square method, respectively.  We find the insignificantly correlations between $\sigma_d$ and black hole mass $\rm M_{BH}$  in NLS1 and BLS1 samples, but significantly positive correlation when combining the two samples which is consistent with \citet{2010ApJ...716L..31A} and \citet{2017ApJ...842...96R}. \citet{2010ApJ...716L..31A} further found the correlation disappeared when eliminate the effect of Eddington ratio. \citet{2022RAA....22a5014W}  showed a weak positive correlation in the 11101 NLS1 galaxies sample.   

\subsubsection{ The dependence of variability with Eddington ratio }         
The Eddington ratio describes the accretion rate of AGNs. It is defined by the ratio of the bolometric luminosity to Eddington luminosity ($\rm \xi = L_{bol}/L_{Edd}$), in which $\rm L_{bol} = 9\times\lambda L_{\lambda}(5100 \AA)$ $\rm erg~s^{-1}$ and $\rm L_{Edd}= 1.3\times10^{38} M_{BH}/M_{\odot}$ $\rm erg~s^{-1}$ \citep{2000ApJ...533..631K}. The relationship between the variability amplitude and Eddington ratio is shown in the upper right panel of Figure 3. The Eddington ratio of NLS1 galaxies is larger than that in BLS1 galaxies which is consistent with  \citet{2017ApJ...842...96R}.  The relationship of variability amplitude and Eddington ratio of NLS1 and BLS1 galaxies shows insignificant anti-correlation, and significant anti-correlation when combining the two samples. \citet{2017ApJ...842...96R} show the uncertainties of black hole mass $\rm M_{BH}$ and Eddington luminosity $L_{\rm Edd}$ may lead to weaken the correlations. The limited range of sample at low redshift domain may also weaken the correlations. 

In order to further study the relation between the variability amplitude and redshift by the finally sample in section 2.1,  we cross match the two samples in Eddington ratio - black hole mass plane (Eddington ratio bin =0.01, black hole bin =0.01),  348 NLS1 galaxies and 348 BLS1 galaxies are found. The two dimension KS test shows the D statistic value of 0.03 and p value 0.999.  The result is shown in Figure 5.  No obviously evolutionary relationship are found between the variability amplitude and redshift. 

\begin{figure}
    \centering
    \includegraphics[width=1\columnwidth]{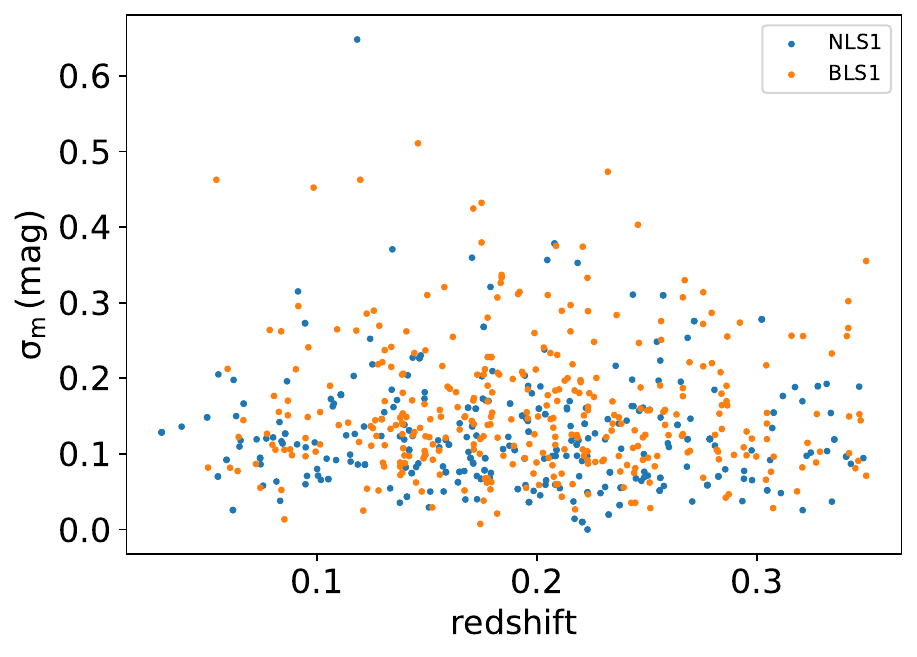}
    \caption{The relation between the variability amplitude and redshift after matching Eddington ratio and black hole mass.  }
    \label{fig:alpha}
\end{figure}


\subsubsection{ The dependence of variability with emission line parameters  }  
We have analyzed the relationship between the variability amplitude and $ R_{\rm Fe \,II}$ as well as $R_{5007}$.   
$R_{5007}$ is defined by the ratio of the flux between total [O III] and total $\rm H\beta$ (broad +narrow) ($ R_{5007} = f[\rm O \,III]_{tot}/f(H\beta)_{tot}$). $ R_{\rm Fe \,II}$ is defined by the ratio of the flux of $\rm Fe \, II$ line to the broad proportion of $\rm H\beta$ line 
($  R_{\rm Fe \,II}=f( \rm Fe \,II(\lambda 4435-4685 $\AA$))/  f(\rm H\beta_{ bro})$). The results are shown in the lower left and right panel of Figure 3, 
wherein no obvious difference are found between the variability amplitude and the emission line parameters between the NLS1 galaxies and BLS1 galaxies sample.  
The results show insignificant correlations between the variability amplitude and $R_{5007} $ as well as $ R_{\rm Fe \,II}$  in NLS1 and BLS1 galaxies sample, significantly positive correlations with $R_{5007}$ and significantly negative correlations with $ R_{\rm Fe \,II}$ when combining the two samples.         

\citet{2010ApJ...716L..31A} show the variability amplitude was significantly negative correlated with $ R_{\rm Fe \,II}$, nevertheless positive correlation with $R_{5007}$.   
 \citet{2017ApJ...842...96R} found the amplitude variability was significantly negative correlation with $R_{\rm Fe \,II}$, which could be explained by the contribution of Fe II emission.  However,  no correlation are found between the variability amplitude and $R_{5007}$. Our results further verify the correlations between the optical variability and emission line parameters $R_{5007}$ and $ R_{\rm Fe \,II}$.

 \begin{table*}
 \caption{Correlation of amplitude of variability ($\sigma_d$) with different AGN parameters. The columns are as follows: (1) band; (2) sample; (3) size of the sample. Columns (4)-(8) note the Spearman correlation coefficient (the $p$-value of no correlation) for the width of $\log \lambda L_{5100}$, $\log M_{\mathrm{BH}}/M_{\odot}$, $\log \lambda_{\mathrm{Edd}}$, $R_{5007}$ and $ R_{\rm Fe \,II}$.}
	\begin{center}
 	\resizebox{\linewidth}{!}{%
     \begin{tabular}{ l l l r r r r r r r}\hline \hline 
       Band & Sample   & Size  &  $\log \lambda L_{5100}$   &  $\log M_{\mathrm{BH}}/M_{\odot}$   &  $\log \lambda_{\mathrm{Edd}}$   &  $R_{5007}$   &  $ R_{\rm Fe \,II}$     \\ 
    (1)  & (2) & (3) & (4) & (5) & (6) & (7) & (8)   \\ \hline
     g 
     &  NLS1      & 2530 &$-$0.008(0.678)    & $+$0.012(0.557)   & $-$0.028(0.147)  & $+$0.021(0.288)   &  $-$0.010(0.620)       \\
     & BLS1       & 2472 & $-$0.001(0.955)   & $-$0.001(0.961)    & $-$0.008(0.702)  & $-$0.002(0.917)    & $-$0.010(0.021)        \\
     & All            & 5002 &  $-$0.005(0.739)    &$+$0.071(5e-08)  & $-$0.121(8e-18)  &  $+$0.054(1e-04)     &   $-$0.043(0.006)       \\           
     \hline 
     r
     &  NLS1      & 2841 & $-$0.037(0.049)     & $-$0.022(0.252)    & $-$0.024(0.196)   &  $+$0.033(0.078)    &  $+$0.019(0.293)      \\   
     & BLS1       & 2798 &  $-$0.020(0.485)    &  $-$0.011(0.551)   &  $-$0.015(0.473)   &  $+$0.013(0.502)    &  $+$0.030(0.223)     \\   
     & All            & 5639 &  $-$0.029(0.025)    &  $+$0.068(3e-07)  &   $-$0.123(1e-20)  & $+$0.067(4e-07)      &  $-$0.011(0.482)          \\   
     \hline         
     i
     &  NLS1      & 2978 & $-$0.005(0.766)     & $+$0.009(0.634)    &  $-$0.016(0.379)     & $+$0.015(0.387)    & $-$0.010(0.571)           \\   
     & BLS1       & 2976 & $-$0.015(0.425)      & $-$0.007(0.713)    &  $+$0.005(0.777)    & $-$0.002(0.896)     & $+$0.001(0.985)        \\   
     & All            &  5954 & $-$0.008(0.523)     & $+$0.086(2e-11)   &  $-$0.120(1e-20)    & $+$0.053(3e-05)    &  $-$0.046(0.001)         \\   
    \hline
     z
     &  NLS1      & 2918 & $-$0.009(0.613)    & $-$0.015(0.412)    & $+$0.012(0.514)      & $+$0.010(0.585)     &-$0.016$(0.381)           \\   
     & BLS1       & 2942 &  $-$0.019(0.302)   &  $-$0.019(0.288)   & $+$0.005(0.773)      &  $-$0.006(0.751)    & $+$0.068(0.004)           \\   
     & All            &  5860 & $-$0.012(0.368)   &  $+$0.083(1e-10)   & $-$0.123(2e-21)       &  $+$0.056(1e-05)    & $-$0.006(0.673)          \\   
     \hline 
     y
     &  NLS1      & 2966 &  $-$0.007(0.668)    &  $-$0.011(0.534)   & $+$0.002(0.922)      & $+$0.011(0.552)     & $-$0.015(0.426)         \\   
     & BLS1       & 2960 &   $-$0.011(0.570)   &  $+$0.001(0.970)  & $-$0.006(0.742)       & $-$0.026(0.158)      & $+$0.019(0.416)          \\   
     & All            & 5926 &  $-$0.010(0.453)    & $+$0.071(5e-08)   & $-$0.071(5e-08)      & $+$0.035(0.008)     & $-$0.037(0.013)      &        \\   
    \hline  \hline
        \end{tabular} } 
        \label{Table:corr1}
        \end{center}
    \end{table*}

\begin{center}    
 \begin{table*}     
 \caption{Correlation of amplitude of variability ($\sigma_d$) in g band with different parameters as Table 1. The values given in columns 4-8 are the Spearman correlation coefficient (the $p$-value of no correlation).}
 	\resizebox{\linewidth}{!}{%
     \begin{tabular}{ l l r r r r r r r}\hline \hline   
       Sample & redshift    & Size  &  $\log \lambda L_{5100}$   &   $\log M_{\mathrm{BH}}/M_{\odot}$   &  $\log \lambda_{\mathrm{Edd}}$   &  $R_{5007}$   &  $\rm R_{Fe \,II}$     \\   \hline       
    NLS1     &                   &  279      & $+$0.043(0.477)      &   $+$0.075(0.212)   &  $-$0.053(0.381)         &  $+$0.008(0.900)     & $-$0.024(0.691)    &   \\
    BLS1     &    0.0-0.1    &   281     & $-$0.003(0.955)       &   $+$0.050(0.403)   &  $-$0.072(0.230)        &  $-$0.017(0.771)     & $-$0.153(0.085)     &        \\  
     All         &                   &   570     & $+$0.002(0.619)      &   $+$0.032(0.026)    &  $-$0.122(0.003)         &  $+$0.032(0.385)      & $-$0.052(0.295)      \\  \hline 
    NLS1     &                  &   930      & $-$0.034(0.288)       &  $+$0.013(0.694)     & $-$0.050(0.124)        &  $+$0.072(0.027)     & $-$0.028(0.385)    &         \\
    BLS1      &    0.1-0.2  &   915      & $+$0.020(0.545)      &  $-$0.008(0.812)      & $+$0.062(0.059)       &  $-$0.032(0.339)      & $-$0.005(0.907)     &        \\
     All          &                 &   1845    & $-$ 0.007(0.761)      &  $+$0.065(0.002)     & $-$0.121(1e-07)         &  $+$0.074(1e-3)      & $-$0.032(0.229)           \\     \hline     
     NLS1     &                 &  1321     & $-$0.002(0.937)       &  $+$0.004(0.894)     & $-$0.009(0.725)        &  $-$0.011(0.702)     & $-$0.048(0.082)     &         \\ 
     BLS1     &    0.2-0.3  &  1276     & $+$0.011(0.702)       & $+$0.016(0.563)      &  $-$0.038(0.139)       &  $+$0.011(0.424)     & $-$0.005(0.878)    &        \\
     All          &                 &  2597     & $-$ 0.002(0.919)      &  $+$0.089(8e-07)     &  $-$0.121(2e-11)       &   $+$0.039(0.043)     & $-$0.048(0.025)          \\           
    \hline \hline       
    \end{tabular} }
     	\label{Table:corr2}
\end{table*}
\end{center}

\subsection{ The relationship between the variability amplitude and the physical parameters in different redshift bins  } 
In order to further investigate the relationship between the variability amplitude and the physical parameters in details, we divide the total sample into three bins ( z = 0-0.1, 0.1-0.2, 0.2-0.35). The correlations in g band are shown in Figure 4. The results of the analysis are given in Table 2. The other results in r, i, z and y bands are listed in the appendix part of Table A1 to A4, respectively.     
 The variability amplitude is significantly positive correlation with black hole mass and $ R_{5007}$, negative correlation with Eddington ratio and $ R_{\rm Fe \,II}$,  insignificant correlation with the luminosity in 5100 \AA.    
 In the three redshift bins, the black hole mass of NLS1 galaxies is smaller than that in BLS1 galaxies. The Eddington ratio of NLS1 galaxies is larger than that in BLS1 galaxies. The variability amplitude between BLS1 galaxies and NLS1 galaxies shows no obvious difference.  \citet{2017ApJ...842...96R} analyzed the correlations in low redshift bins (z<0.4) which is consistent with our results.

\subsection{ The relation of variability amplitude and radio loudness in the radio subsample  } 
In order to investigate the relationship between the variability amplitude and radio loudness, we construct a radio subsample by crossing match with the Faint Images of the Radio Sky at Twenty centimeters (FIRST) survey with the search radius five arcseconds. We utilize 3192 BLS1 sample and 3194 NLS1 sample to cross match with the FIRST survey, 155 BLS1 galaxies and 188 NLS1 galaxies are left. We further consider the data sets number is larger than 5 and the timespan is longer than 2 years, thus 149 BLS1 galaxies and 177 NLS1 galaxies are left in i band. The similar method are used to deal with the data sets in g, r, z and y bands, respectively. Finally, we obtain 124 BLS1 galaxies and 146 NLS1 galaxies in g band, 142 BLS1 galaxies and 170 NLS1 galaxies in r band, 147 BLS1 galaxies and 179 NLS1 galaxies in z band, 147 BLS1 galaxies and 174 NLS1 galaxies in y band, respectively. The radio loudness is defined as the ratio between the flux in radio 1.4 GHz and the optical flux in g band ($\rm R= f_{1.4 GHz/}/f_g$). The results are listed in Figure 6, wherein no significant correlations were displayed. The results also show no significant difference between the NLS1 galaxies and BLS1 galaxies.   

\citet{Rakshit2017}  investigated the relationship between the variability amplitude and radio loudness by the optical V-band light curves from the CRTS, wherein it showed significantly positive correlation. They considered that some contribution of the optical emission in both samples may be originated from the relativistic jet, hence the radio loud objects show more variable than radio quiet objects. It also explains the positive correlation between the variability amplitude and radio loudness. However, our results display no significant correlation in Figure 6. The time baseline of Pan-STARRS in our sample is $2.0\sim 5.6$ years which is shorter than $5\sim6$ years in CRTS. The variability amplitude is related to the observed time baseline and cadence, and thus influence the measuring result. Therefore, more longer time baseline should be investigated the results again in the future.

\begin{figure*} 
\begin{subfigure}{.33\textwidth}
  \centering   
  \includegraphics[width=.98\linewidth]{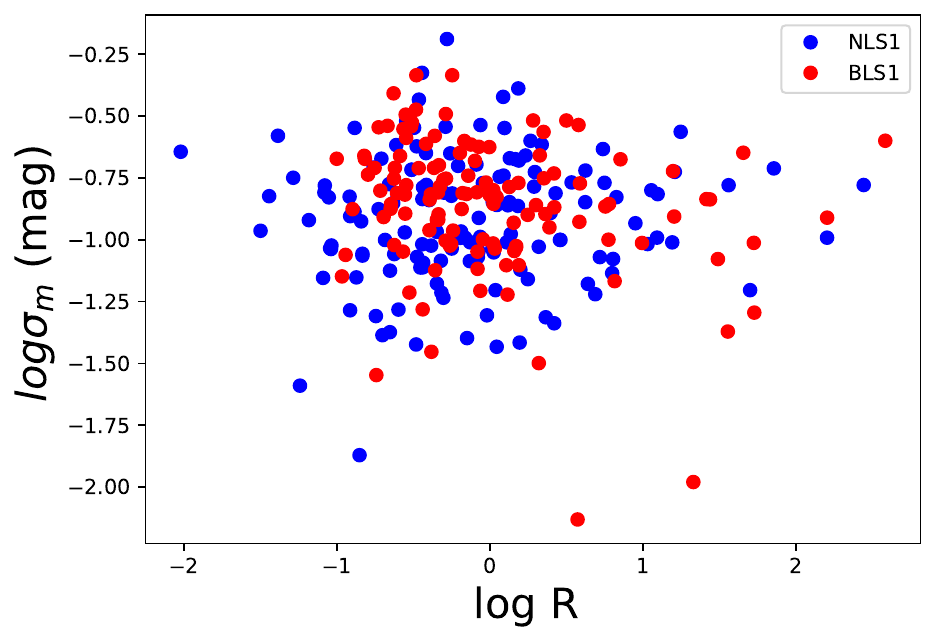}  
  \caption{ The results in g band }         
  \label{fig:sub-first}
\end{subfigure}       
\begin{subfigure}{.33\textwidth}     
  \centering
  \includegraphics[width=.98\linewidth]{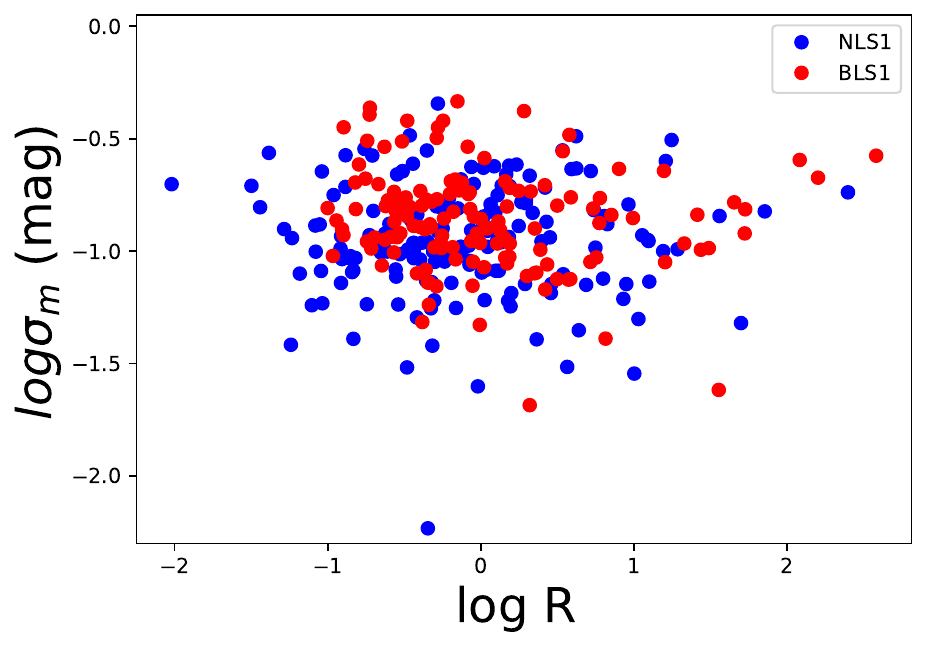}  
  \caption{ The results in r band }
  \label{fig:sub-second}
\end{subfigure}
\hfill
\begin{subfigure}{.33\textwidth}
  \centering
  \includegraphics[width=.98\linewidth]{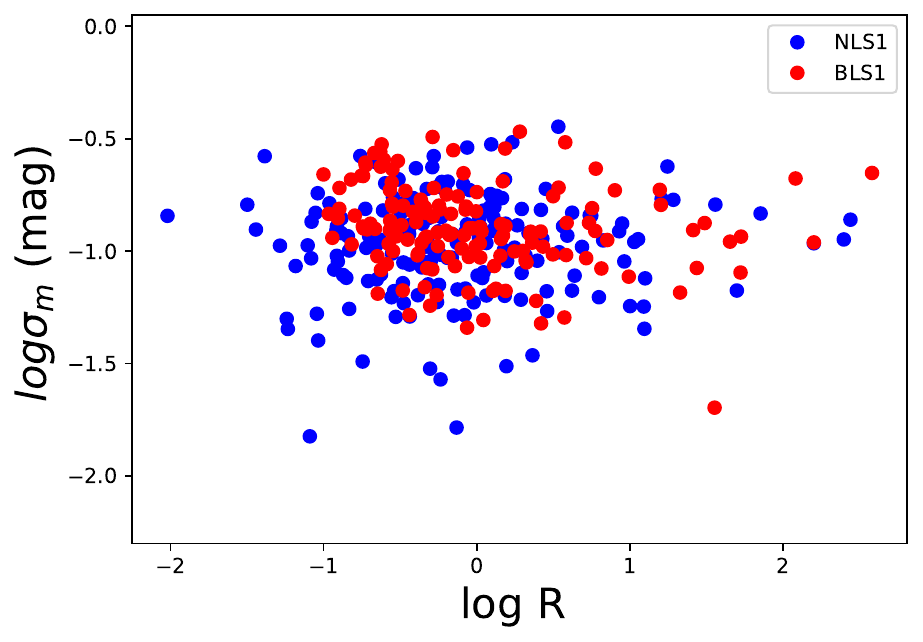}  
  \caption{ The results in i band }
  \label{fig:sub-second}
\end{subfigure}
\hfill
\begin{subfigure}{.33\textwidth}
  \centering
  \includegraphics[width=.98\linewidth]{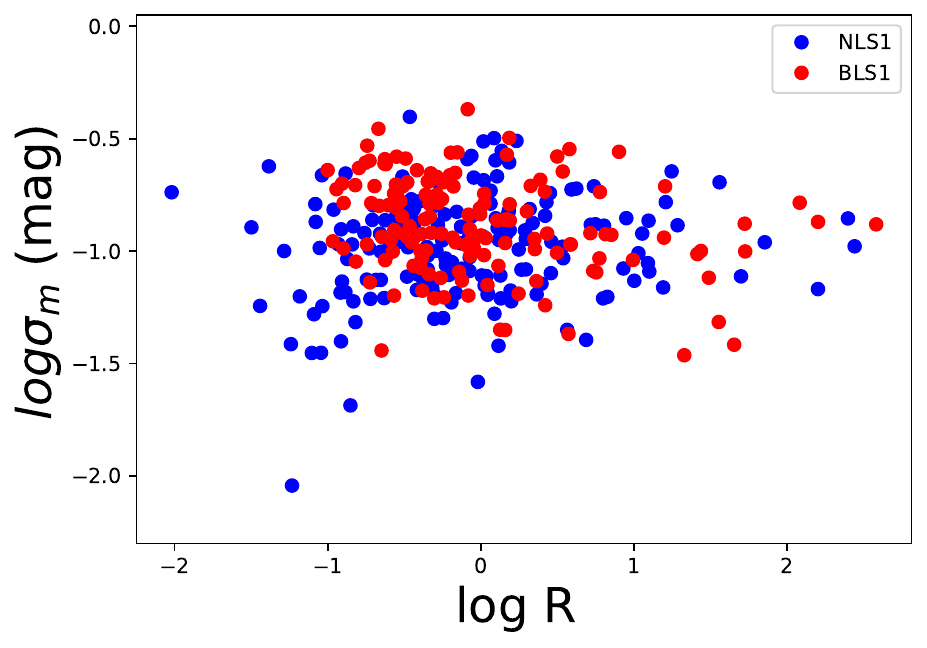}  
  \caption{  The results in z band}
  \label{fig:sub-second}
\end{subfigure}
\begin{subfigure}{.33\textwidth}
  \centering
  \includegraphics[width=.98\linewidth]{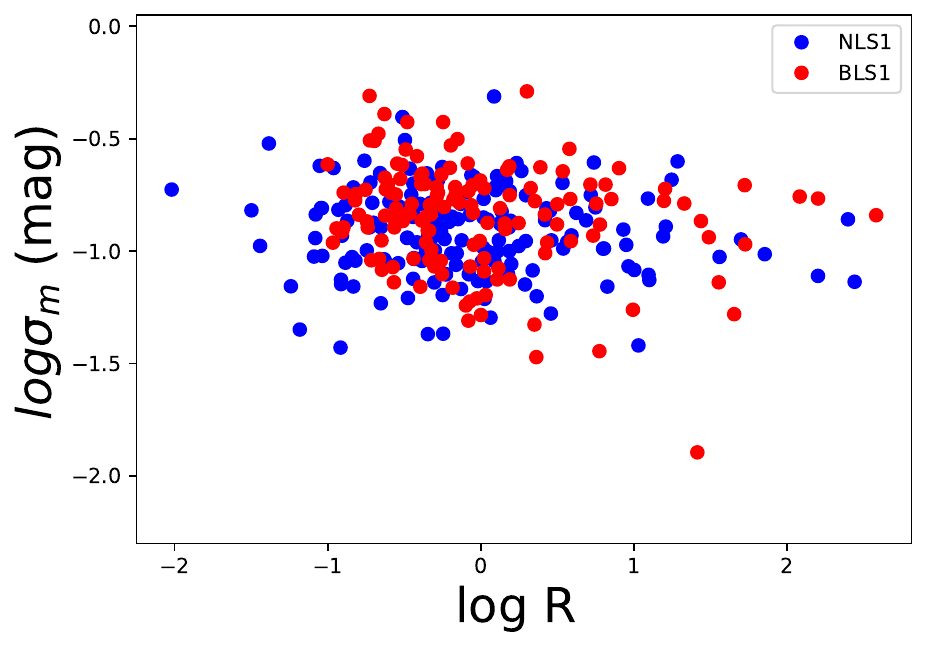}  
  \caption{ The results in y band  }
  \label{fig:sub-second}   
\end{subfigure}
\caption{  The relation between variability amplitude and radio loudness of BLS1 and NLS1 galaxies in g, r, i, z and y bands, respectively.  }
\label{fig:fig}
\end{figure*}

\section{Summary}
\label{sec:summary}
By means of the data sets of the Pan-STARRS, we investigate the relationship between the variability amplitude and the luminosity at 5100 \AA,  black hole mass, Eddington ratio, $\rm R_{Fe \,II}$ as well as $\rm R_{5007}$ of the BLS1 and NLS1 galaxies sample in g,r,i,z and y bands. We also analyze the similarities and differences of the variability characteristics between BLS1 galaxies and NLS1 galaxies. The results are listed as follows.  

(1). The median variability amplitude is 0.142 mag for BLS1 galaxies and 0.119 mag for NLS1 galaxies in g band,  0.137 mag for BLS1 galaxies and 0.114 mag for NLS1 galaxies in r band, 0.130 mag for BLS1 galaxies and 0.112 mag for NLS1 galaxies in i band,  0.125 mag for BLS1 galaxies and 0.103 mag for NLS1 galaxies in z band,  0.132 mag for NLS1 galxies and 0.115 mag for NLS1 galaxies in y band, respectively.  The cumulative probability distribution of variability amplitude of NLS1 galaxies is lower than that in BLS1 galaxies.  

(2). We analyze the dependence of the variability amplitude with the luminosity at 5100 \AA, black hole mass, Eddington ratio, $\rm R_{Fe \,II}$ and $\rm R_{5007}$, respectively.  We find the significantly negative correlations between the variability amplitude and Eddington ratio, insignificant correlations with luminosity at 5100 \AA. The results also show significantly positive correlations with the black hole mass and $\rm R_{5007}$, significantly negative correlations with $\rm R_{Fe \,II}$. The results are consistent with \citet{2017ApJ...842...96R} in low redshift bins (z<0.4) and \citet{2010ApJ...716L..31A}.  

(3). We further find that the $\rm M_{BH}$ of NLS1 galaxies is obviously smaller than BLS1 galaxies. The Eddington ratio of NLS1 galxies is noticeably larger than BLS1 galaxies. The results may originate from the structure of discold BLR, rather than isotropic geometry.    

(4). The relationship between the variability amplitude and the radio loudness is investigated for 155 BLS1 galaxies and 188 NLS1 galaxies. No significant  correlations are found in our results.

\section*{Acknowledgement}
The authors would like to thank the referee for providing valuable suggestions and comments, which helped in improving the manuscript significantly.
  H.T.W is supported by the Hebei Natural Science Foundation of China (Grant No. A2022408002) and the Fundamental Research Funds for the Universities in Hebei Province (Grant No. JYQ202003).  N.D are grateful for the financial support from the National Natural Science Foundation of China (No. 12103022) and the Special Basic Cooperative Research Programs of Yunnan Provincial 
  Undergraduate Universities  Association (No.202101BA070001-043). 
  
 The Pan-STARRS1 Surveys (PS1) and the PS1 public science archive have been made possible through contributions by the Institute for Astronomy, the University of Hawaii, the Pan-STARRS Project Office, the Max-Planck Society and its participating institutes, the Max Planck Institute for Astronomy, Heidelberg and the Max Planck Institute for Extraterrestrial Physics, Garching, The Johns Hopkins University, Durham University, the University of Edinburgh, the Queen's University Belfast, the Harvard-Smithsonian Center for Astrophysics, the Las Cumbres Observatory Global Telescope Network Incorporated, the National Central University of Taiwan, the Space Telescope Science Institute, the National Aeronautics and Space Administration under Grant No. NNX08AR22G issued through the Planetary Science Division of the NASA Science Mission Directorate, the National Science Foundation Grant No. AST-1238877, the University of Maryland, Eotvos Lorand University (ELTE), the Los Alamos National Laboratory, and the Gordon and Betty Moore Foundation.



\bibliographystyle{mnras}
\bibliography{references}   

\begin{thebibliography}{}
\makeatletter
\relax
\def\mn@urlcharsother{\let\do\@makeother \do\$\do\&\do\#\do\^\do\_\do\%\do\~}
\def\mn@doi{\begingroup\mn@urlcharsother \@ifnextchar [ {\mn@doi@}
  {\mn@doi@[]}}
\def\mn@doi@[#1]#2{\def\@tempa{#1}\ifx\@tempa\@empty \href
  {http://dx.doi.org/#2} {doi:#2}\else \href {http://dx.doi.org/#2} {#1}\fi
  \endgroup}
\def\mn@eprint#1#2{\mn@eprint@#1:#2::\@nil}
\def\mn@eprint@arXiv#1{\href {http://arxiv.org/abs/#1} {{\tt arXiv:#1}}}
\def\mn@eprint@dblp#1{\href {http://dblp.uni-trier.de/rec/bibtex/#1.xml}
  {dblp:#1}}
\def\mn@eprint@#1:#2:#3:#4\@nil{\def\@tempa {#1}\def\@tempb {#2}\def\@tempc
  {#3}\ifx \@tempc \@empty \let \@tempc \@tempb \let \@tempb \@tempa \fi \ifx
  \@tempb \@empty \def\@tempb {arXiv}\fi \@ifundefined
  {mn@eprint@\@tempb}{\@tempb:\@tempc}{\expandafter \expandafter \csname
  mn@eprint@\@tempb\endcsname \expandafter{\@tempc}}}

\bibitem[\protect\citeauthoryear{{Ai}, {Yuan}, {Zhou}, {Wang}, {Dong}, {Wang}
  \& {Lu}}{{Ai} et~al.}{2010}]{2010ApJ...716L..31A}
{Ai} Y.~L.,  {Yuan} W.,  {Zhou} H.~Y.,  {Wang} T.~G.,  {Dong} X.~B.,  {Wang}
  J.~G.,   {Lu} H.~L.,  2010, \mn@doi [\apjl] {10.1088/2041-8205/716/1/L31},
  \href {https://ui.adsabs.harvard.edu/abs/2010ApJ...716L..31A} {716, L31}

\bibitem[\protect\citeauthoryear{{Ai}, {Yuan}, {Zhou}, {Wang}, {Dong}, {Wang}
  \& {Lu}}{{Ai} et~al.}{2013}]{2013AJ....145...90A}
{Ai} Y.~L.,  {Yuan} W.,  {Zhou} H.,  {Wang} T.~G.,  {Dong} X.~B.,  {Wang}
  J.~G.,   {Lu} H.~L.,  2013, \mn@doi [\aj] {10.1088/0004-6256/145/4/90}, \href
  {https://ui.adsabs.harvard.edu/abs/2013AJ....145...90A} {145, 90}

\bibitem[\protect\citeauthoryear{{Aretxaga}, {Cid Fernandes}  \&
  {Terlevich}}{{Aretxaga} et~al.}{1997}]{1997MNRAS.286..271A}
{Aretxaga} I.,  {Cid Fernandes} R.,   {Terlevich} R.~J.,  1997, \mn@doi
  [\mnras] {10.1093/mnras/286.2.271}, \href
  {https://ui.adsabs.harvard.edu/abs/1997MNRAS.286..271A} {286, 271}

\bibitem[\protect\citeauthoryear{{Baldi}, {Capetti}, {Robinson}, {Laor}  \&
  {Behar}}{{Baldi} et~al.}{2016}]{2016MNRAS.458L..69B}
{Baldi} R.~D.,  {Capetti} A.,  {Robinson} A.,  {Laor} A.,   {Behar} E.,  2016,
  \mn@doi [\mnras] {10.1093/mnrasl/slw019}, \href
  {https://ui.adsabs.harvard.edu/abs/2016MNRAS.458L..69B} {458, L69}

\bibitem[\protect\citeauthoryear{{Boller}, {Brandt}  \& {Fink}}{{Boller}
  et~al.}{1996}]{1996A&A...305...53B}
{Boller} T.,  {Brandt} W.~N.,   {Fink} H.,  1996, \mn@doi [\aap]
  {10.48550/arXiv.astro-ph/9504093}, \href
  {https://ui.adsabs.harvard.edu/abs/1996A&A...305...53B} {305, 53}

\bibitem[\protect\citeauthoryear{{Boroson} \& {Green}}{{Boroson} \&
  {Green}}{1992}]{1992ApJS...80..109B}
{Boroson} T.~A.,  {Green} R.~F.,  1992, \mn@doi [\apjs] {10.1086/191661}, \href
  {https://ui.adsabs.harvard.edu/abs/1992ApJS...80..109B} {80, 109}

\bibitem[\protect\citeauthoryear{{Calderone}, {Ghisellini}, {Colpi}  \&
  {Dotti}}{{Calderone} et~al.}{2013}]{2013MNRAS.431..210C}
{Calderone} G.,  {Ghisellini} G.,  {Colpi} M.,   {Dotti} M.,  2013, \mn@doi
  [\mnras] {10.1093/mnras/stt157}, \href
  {https://ui.adsabs.harvard.edu/abs/2013MNRAS.431..210C} {431, 210}

\bibitem[\protect\citeauthoryear{{Cid Fernandes}, {Sodr{\'e}}  \& {Vieira da
  Silva}}{{Cid Fernandes} et~al.}{2000}]{2000ApJ...544..123C}
{Cid Fernandes} R.,  {Sodr{\'e}} L. J.,   {Vieira da Silva} L. J.,  2000,
  \mn@doi [\apj] {10.1086/317207}, \href
  {https://ui.adsabs.harvard.edu/abs/2000ApJ...544..123C} {544, 123}

\bibitem[\protect\citeauthoryear{{Decarli}, {Dotti}, {Fontana}  \&
  {Haardt}}{{Decarli} et~al.}{2008}]{Decarli2008}
{Decarli} R.,  {Dotti} M.,  {Fontana} M.,   {Haardt} F.,  2008, \mn@doi
  [\mnras] {10.1111/j.1745-3933.2008.00451.x}, \href
  {https://ui.adsabs.harvard.edu/abs/2008MNRAS.386L..15D} {386, L15}

\bibitem[\protect\citeauthoryear{{Fasano} \& {Franceschini}}{{Fasano} \&
  {Franceschini}}{1987}]{Fasano1987}
{Fasano} G.,  {Franceschini} A.,  1987, \mn@doi [\mnras]
  {10.1093/mnras/225.1.155}, \href
  {https://ui.adsabs.harvard.edu/abs/1987MNRAS.225..155F} {225, 155}

\bibitem[\protect\citeauthoryear{{Gaskell}}{{Gaskell}}{1984}]{1984ApL....24...43G}
{Gaskell} C.~M.,  1984, \aplett, \href
  {https://ui.adsabs.harvard.edu/abs/1984ApL....24...43G} {24, 43}

\bibitem[\protect\citeauthoryear{{Gaskell}}{{Gaskell}}{2008}]{2008RMxAC..32....1G}
{Gaskell} C.~M.,  2008, in Revista Mexicana de Astronomia y Astrofisica
  Conference Series. pp 1--11 (\mn@eprint {arXiv} {0711.2113}),
  \mn@doi{10.48550/arXiv.0711.2113}

\bibitem[\protect\citeauthoryear{{Gaskell} \& {Kormendy}}{{Gaskell} \&
  {Kormendy}}{2009}]{2009ASPC..419..388G}
{Gaskell} C.~M.,  {Kormendy} J.,  2009, in {Jogee} S.,  {Marinova} I.,  {Hao}
  L.,   {Blanc} G.~A.,  eds,  Astronomical Society of the Pacific Conference
  Series Vol. 419, Galaxy Evolution: Emerging Insights and Future Challenges.
  p.~388 (\mn@eprint {arXiv} {0907.1652}), \mn@doi{10.48550/arXiv.0907.1652}

\bibitem[\protect\citeauthoryear{{Gaskell} \& {Peterson}}{{Gaskell} \&
  {Peterson}}{1987}]{1987ApJS...65....1G}
{Gaskell} C.~M.,  {Peterson} B.~M.,  1987, \mn@doi [\apjs] {10.1086/191216},
  \href {https://ui.adsabs.harvard.edu/abs/1987ApJS...65....1G} {65, 1}

\bibitem[\protect\citeauthoryear{{Guo}, {Wang}, {Cai}  \& {Sun}}{{Guo}
  et~al.}{2017}]{2017ApJ...847..132G}
{Guo} H.,  {Wang} J.,  {Cai} Z.,   {Sun} M.,  2017, \mn@doi [\apj]
  {10.3847/1538-4357/aa8d71}, \href
  {https://ui.adsabs.harvard.edu/abs/2017ApJ...847..132G} {847, 132}

\bibitem[\protect\citeauthoryear{{Hawkins}}{{Hawkins}}{2000}]{2000A&AS..143..465H}
{Hawkins} M.~R.~S.,  2000, \mn@doi [\aaps] {10.1051/aas:2000190}, \href
  {https://ui.adsabs.harvard.edu/abs/2000A&AS..143..465H} {143, 465}

\bibitem[\protect\citeauthoryear{{Ivezic} et~al.,}{{Ivezic}
  et~al.}{2004}]{2004IAUS..222..525I}
{Ivezic} {\v{Z}}.,  et~al., 2004, in {Storchi-Bergmann} T.,  {Ho} L.~C.,
  {Schmitt} H.~R.,  eds,  Vol. 222, The Interplay Among Black Holes, Stars and
  ISM in Galactic Nuclei. pp 525--526 (\mn@eprint {arXiv} {astro-ph/0404487}),
  \mn@doi{10.1017/S1743921304003126}

\bibitem[\protect\citeauthoryear{{Kang}, {Wang}, {Cai}  \& {Ren}}{{Kang}
  et~al.}{2021}]{2021ApJ...911..148K}
{Kang} W.-Y.,  {Wang} J.-X.,  {Cai} Z.-Y.,   {Ren} W.-K.,  2021, \mn@doi [\apj]
  {10.3847/1538-4357/abeb69}, \href
  {https://ui.adsabs.harvard.edu/abs/2021ApJ...911..148K} {911, 148}

\bibitem[\protect\citeauthoryear{{Kaspi}, {Smith}, {Netzer}, {Maoz}, {Jannuzi}
  \& {Giveon}}{{Kaspi} et~al.}{2000}]{2000ApJ...533..631K}
{Kaspi} S.,  {Smith} P.~S.,  {Netzer} H.,  {Maoz} D.,  {Jannuzi} B.~T.,
  {Giveon} U.,  2000, \mn@doi [\apj] {10.1086/308704}, \href
  {https://ui.adsabs.harvard.edu/abs/2000ApJ...533..631K} {533, 631}

\bibitem[\protect\citeauthoryear{{Kawaguchi}, {Mineshige}, {Umemura}  \&
  {Turner}}{{Kawaguchi} et~al.}{1998}]{1998ApJ...504..671K}
{Kawaguchi} T.,  {Mineshige} S.,  {Umemura} M.,   {Turner} E.~L.,  1998,
  \mn@doi [\apj] {10.1086/306105}, \href
  {https://ui.adsabs.harvard.edu/abs/1998ApJ...504..671K} {504, 671}

\bibitem[\protect\citeauthoryear{{Kelly}, {Bechtold}  \&
  {Siemiginowska}}{{Kelly} et~al.}{2009}]{2009ApJ...698..895K}
{Kelly} B.~C.,  {Bechtold} J.,   {Siemiginowska} A.,  2009, \mn@doi [\apj]
  {10.1088/0004-637X/698/1/895}, \href
  {https://ui.adsabs.harvard.edu/abs/2009ApJ...698..895K} {698, 895}

\bibitem[\protect\citeauthoryear{{Klimek}, {Gaskell}  \& {Hedrick}}{{Klimek}
  et~al.}{2004}]{2004ApJ...609...69K}
{Klimek} E.~S.,  {Gaskell} C.~M.,   {Hedrick} C.~H.,  2004, \mn@doi [\apj]
  {10.1086/420809}, \href
  {https://ui.adsabs.harvard.edu/abs/2004ApJ...609...69K} {609, 69}

\bibitem[\protect\citeauthoryear{{Li} \& {Cao}}{{Li} \&
  {Cao}}{2008}]{2008MNRAS.387L..41L}
{Li} S.-L.,  {Cao} X.,  2008, \mn@doi [\mnras]
  {10.1111/j.1745-3933.2008.00480.x}, \href
  {https://ui.adsabs.harvard.edu/abs/2008MNRAS.387L..41L} {387, L41}

\bibitem[\protect\citeauthoryear{{Liu}, {Yang}, {Supriyanto}  \& {Zhang}}{{Liu}
  et~al.}{2016}]{2016IJAA....6..166L}
{Liu} X.,  {Yang} P.,  {Supriyanto} R.,   {Zhang} Z.,  2016, \mn@doi
  [International Journal of Astronomy and Astrophysics]
  {10.4236/ijaa.2016.62014}, \href
  {https://ui.adsabs.harvard.edu/abs/2016IJAA....6..166L} {6, 166}

\bibitem[\protect\citeauthoryear{{Liu}, {Liu}, {Dong}, {Zhou}, {Wang}, {Lu}  \&
  {Yuan}}{{Liu} et~al.}{2019}]{LiuHY2019ApJS}
{Liu} H.-Y.,  {Liu} W.-J.,  {Dong} X.-B.,  {Zhou} H.,  {Wang} T.,  {Lu} H.,
  {Yuan} W.,  2019, \mn@doi [\apjs] {10.3847/1538-4365/ab298b}, \href
  {https://ui.adsabs.harvard.edu/abs/2019ApJS..243...21L} {243, 21}

\bibitem[\protect\citeauthoryear{{Osterbrock}}{{Osterbrock}}{1977}]{1977ApJ...215..733O}
{Osterbrock} D.~E.,  1977, \mn@doi [\apj] {10.1086/155407}, \href
  {https://ui.adsabs.harvard.edu/abs/1977ApJ...215..733O} {215, 733}

\bibitem[\protect\citeauthoryear{{Osterbrock} \& {Pogge}}{{Osterbrock} \&
  {Pogge}}{1985}]{1985ApJ...297..166O}
{Osterbrock} D.~E.,  {Pogge} R.~W.,  1985, \mn@doi [\apj] {10.1086/163513},
  \href {https://ui.adsabs.harvard.edu/abs/1985ApJ...297..166O} {297, 166}

\bibitem[\protect\citeauthoryear{{Osterbrock}, {Koski}  \&
  {Phillips}}{{Osterbrock} et~al.}{1975}]{1975ApJ...197L..41O}
{Osterbrock} D.~E.,  {Koski} A.~T.,   {Phillips} M.~M.,  1975, \mn@doi [\apjl]
  {10.1086/181772}, \href
  {https://ui.adsabs.harvard.edu/abs/1975ApJ...197L..41O} {197, L41}

\bibitem[\protect\citeauthoryear{{Osterbrock}, {Koski}  \&
  {Phillips}}{{Osterbrock} et~al.}{1976}]{1976ApJ...206..898O}
{Osterbrock} D.~E.,  {Koski} A.~T.,   {Phillips} M.~M.,  1976, \mn@doi [\apj]
  {10.1086/154454}, \href
  {https://ui.adsabs.harvard.edu/abs/1976ApJ...206..898O} {206, 898}

\bibitem[\protect\citeauthoryear{{Panda}, {Marziani}  \& {Czerny}}{{Panda}
  et~al.}{2019}]{2019ApJ...882...79P}
{Panda} S.,  {Marziani} P.,   {Czerny} B.,  2019, \mn@doi [\apj]
  {10.3847/1538-4357/ab3292}, \href
  {https://ui.adsabs.harvard.edu/abs/2019ApJ...882...79P} {882, 79}

\bibitem[\protect\citeauthoryear{{Peacock}}{{Peacock}}{1983}]{Peacock1983}
{Peacock} J.~A.,  1983, \mn@doi [\mnras] {10.1093/mnras/202.3.615}, \href
  {https://ui.adsabs.harvard.edu/abs/1983MNRAS.202..615P} {202, 615}

\bibitem[\protect\citeauthoryear{{Puchnarewicz} et~al.,}{{Puchnarewicz}
  et~al.}{1992}]{1992MNRAS.256..589P}
{Puchnarewicz} E.~M.,  et~al., 1992, \mn@doi [\mnras]
  {10.1093/mnras/256.3.589}, \href
  {https://ui.adsabs.harvard.edu/abs/1992MNRAS.256..589P} {256, 589}

\bibitem[\protect\citeauthoryear{{Rakshit} \& {Stalin}}{{Rakshit} \&
  {Stalin}}{2017}]{2017ApJ...842...96R}
{Rakshit} S.,  {Stalin} C.~S.,  2017, \mn@doi [\apj]
  {10.3847/1538-4357/aa72f4}, \href
  {https://ui.adsabs.harvard.edu/abs/2017ApJ...842...96R} {842, 96}

\bibitem[\protect\citeauthoryear{{Rakshit}, {Stalin}, {Chand}  \&
  {Zhang}}{{Rakshit} et~al.}{2017}]{Rakshit2017}
{Rakshit} S.,  {Stalin} C.~S.,  {Chand} H.,   {Zhang} X.-G.,  2017, \mn@doi
  [\apjs] {10.3847/1538-4365/aa6971}, \href
  {https://ui.adsabs.harvard.edu/abs/2017ApJS..229...39R} {229, 39}

\bibitem[\protect\citeauthoryear{{Rakshit}, {Johnson}, {Stalin}, {Gandhi}  \&
  {Hoenig}}{{Rakshit} et~al.}{2019}]{2019MNRAS.483.2362R}
{Rakshit} S.,  {Johnson} A.,  {Stalin} C.~S.,  {Gandhi} P.,   {Hoenig} S.,
  2019, \mn@doi [\mnras] {10.1093/mnras/sty3261}, \href
  {https://ui.adsabs.harvard.edu/abs/2019MNRAS.483.2362R} {483, 2362}

\bibitem[\protect\citeauthoryear{{Sesar} et~al.,}{{Sesar}
  et~al.}{2007}]{2007AJ....134.2236S}
{Sesar} B.,  et~al., 2007, \mn@doi [\aj] {10.1086/521819}, \href
  {https://ui.adsabs.harvard.edu/abs/2007AJ....134.2236S} {134, 2236}

\bibitem[\protect\citeauthoryear{{Shen} \& {Ho}}{{Shen} \&
  {Ho}}{2014}]{2014Natur.513..210S}
{Shen} Y.,  {Ho} L.~C.,  2014, \mn@doi [\nat] {10.1038/nature13712}, \href
  {https://ui.adsabs.harvard.edu/abs/2014Natur.513..210S} {513, 210}

\bibitem[\protect\citeauthoryear{{Shen} et~al.,}{{Shen}
  et~al.}{2011}]{2011ApJS..194...45S}
{Shen} Y.,  et~al., 2011, \mn@doi [\apjs] {10.1088/0067-0049/194/2/45}, \href
  {https://ui.adsabs.harvard.edu/abs/2011ApJS..194...45S} {194, 45}

\bibitem[\protect\citeauthoryear{{Sulentic}, {Zwitter}, {Marziani}  \&
  {Dultzin-Hacyan}}{{Sulentic} et~al.}{2000}]{2000ApJ...536L...5S}
{Sulentic} J.~W.,  {Zwitter} T.,  {Marziani} P.,   {Dultzin-Hacyan} D.,  2000,
  \mn@doi [\apjl] {10.1086/312717}, \href
  {https://ui.adsabs.harvard.edu/abs/2000ApJ...536L...5S} {536, L5}

\bibitem[\protect\citeauthoryear{{Sun} \& {Shen}}{{Sun} \&
  {Shen}}{2015}]{2015ApJ...804L..15S}
{Sun} J.,  {Shen} Y.,  2015, \mn@doi [\apjl] {10.1088/2041-8205/804/1/L15},
  \href {https://ui.adsabs.harvard.edu/abs/2015ApJ...804L..15S} {804, L15}

\bibitem[\protect\citeauthoryear{{Sun}, {Xue}, {Wang}, {Cai}  \& {Guo}}{{Sun}
  et~al.}{2018}]{2018ApJ...866...74S}
{Sun} M.,  {Xue} Y.,  {Wang} J.,  {Cai} Z.,   {Guo} H.,  2018, \mn@doi [\apj]
  {10.3847/1538-4357/aae208}, \href
  {https://ui.adsabs.harvard.edu/abs/2018ApJ...866...74S} {866, 74}

\bibitem[\protect\citeauthoryear{{Ulrich}, {Maraschi}  \& {Urry}}{{Ulrich}
  et~al.}{1997}]{1997ARA&A..35..445U}
{Ulrich} M.-H.,  {Maraschi} L.,   {Urry} C.~M.,  1997, \mn@doi [\araa]
  {10.1146/annurev.astro.35.1.445}, \href
  {https://ui.adsabs.harvard.edu/abs/1997ARA&A..35..445U} {35, 445}

\bibitem[\protect\citeauthoryear{{Vanden Berk} et~al.,}{{Vanden Berk}
  et~al.}{2004}]{2004ApJ...601..692V}
{Vanden Berk} D.~E.,  et~al., 2004, \mn@doi [\apj] {10.1086/380563}, \href
  {https://ui.adsabs.harvard.edu/abs/2004ApJ...601..692V} {601, 692}

\bibitem[\protect\citeauthoryear{{V{\'e}ron-Cetty}, {V{\'e}ron}  \&
  {Gon{\c{c}}alves}}{{V{\'e}ron-Cetty} et~al.}{2001}]{2001A&A...372..730V}
{V{\'e}ron-Cetty} M.~P.,  {V{\'e}ron} P.,   {Gon{\c{c}}alves} A.~C.,  2001,
  \mn@doi [\aap] {10.1051/0004-6361:20010489}, \href
  {https://ui.adsabs.harvard.edu/abs/2001A&A...372..730V} {372, 730}

\bibitem[\protect\citeauthoryear{{Wang}, {Su}, {Ge}, {Chen}  \& {Yu}}{{Wang}
  et~al.}{2022}]{2022RAA....22a5014W}
{Wang} H.-T.,  {Su} Y.-P.,  {Ge} X.,  {Chen} Y.-Y.,   {Yu} X.-L.,  2022,
  \mn@doi [Research in Astronomy and Astrophysics] {10.1088/1674-4527/ac3895},
  \href {https://ui.adsabs.harvard.edu/abs/2022RAA....22a5014W} {22, 015014}

\bibitem[\protect\citeauthoryear{{Wilkes}, {Elvis}  \& {McHardy}}{{Wilkes}
  et~al.}{1987}]{1987ApJ...321L..23W}
{Wilkes} B.~J.,  {Elvis} M.,   {McHardy} I.,  1987, \mn@doi [\apjl]
  {10.1086/184999}, \href
  {https://ui.adsabs.harvard.edu/abs/1987ApJ...321L..23W} {321, L23}

\makeatother
\end{thebibliography}

\appendix
\section{Additional results}

\begin{center}    
 \begin{table*}     
 \caption{Correlation of amplitude of variability ($\sigma_d$) in r band with different parameters as Table 1. The values given in columns 4-8 are the Spearman correlation coefficient (the $p$-value of no correlation).}
 	\resizebox{\linewidth}{!}{%
     \begin{tabular}{ l l r r r r r r r}\hline \hline   
       Sample & redshift    & Size  &  $\log \lambda L_{5100}$   &   $\log M_{\mathrm{BH}}/M_{\odot}$   &  $\log \lambda_{\mathrm{Edd}}$   &  $R_{5007}$   &  $\rm R_{Fe \,II}$    \\   \hline       
    NLS1     &                  &   310     &  $-$0.023(0.678)      &   $+$0.012(0.823)    &  $+$0.014(0.804)       &  $-$0.037(0.511)     & $+$0.005(0.933)       &   \\
    BLS1     &    0.0-0.1   &    303     &  $-$0.031(0.589)      &   $+$0.058(0.308)    &  $-$0.009(0.873)       &  $+$0.008(0.883)     &  $-$0.202(0.018)     &        \\  
     All         &                   &    613     &  $-$0.032(0.428)      &   $-$0.088(0.028)     &  $-$0.083(0.039)        &  $+$0.025(0.533)   &  $-$0.091(0.053)      &    \\  \hline 
  NLS1       &                  &   1049     & $-$0.089(0.003)      &   $-$0.054(0.078)    &  $-$0.047(0.126)        &  $+$0.058(0.058)    &  $+$0.055(0.076)    &   \\
    BLS1     &   0.0-0.1    &   1036      &  $-$0.039(0.201)      &   $-$0.037(0.224)    &  $+$0.033(0.283)      &  $+$0.007(0.818)     & $+$0.034(0.418)    &        \\  
     All         &                  &    2085     &  $-$0.063(0.003)      &   $+$0.046(0.033)   &  $-$0.121(3e-08)        &  $+$0.086(8e-05)    & $+$0.005(0.833)    &    \\  \hline 
 NLS1        &                 &   1130      &  $-$0.037(0.146)      &   $-$0.037(0.181)    &  $+$0.022(0.456)        &  $+$0.034(0.186)    & $-$0.009(0.924)    &   \\
    BLS1     &   0.2-0.3    &  1121       &  $-$0.032(0.211)      &   $-$0.001(0.963)    &  $-$0.045(0.130)        &  $+$0.018(0.487)    & $+$0.058(0.071)     &        \\  
     All         &                  &   2251      &  $-$0.035(0.058)      &   $+$0.087(2e-06)    &  $-$0.143(8e-12)      &  $+$0.068(2e-04)    & $-$0.008(0.679)     &    \\   
    \hline \hline       
    \end{tabular} }
     	\label{Table:corr2}
\end{table*}
\end{center}

\begin{center}    
 \begin{table*}     
 \caption{Correlation of amplitude of variability ($\sigma_d$) in i band with different parameters as Table 1. The values given in columns 4-8 are the Spearman correlation coefficient (the $p$-value of no correlation).}
 	\resizebox{\linewidth}{!}{%
     \begin{tabular}{ l l r r r r r r r}\hline \hline       
       Sample & redshift    & Size  &  $\log \lambda L_{5100}$   &   $\log M_{\mathrm{BH}}/M_{\odot}$   &  $\log \lambda_{\mathrm{Edd}}$   &  $R_{5007}$   &  $\rm R_{Fe \,II}$     \\   \hline       
    NLS1     &                   &   295    &  $+$0.027(0.625)      &   $-$0.035(0.535)     &  $+$0.014(0.804)        &  $+$0.044(0.429)      & $+$0.002(0.672)        \\
    BLS1     &   0.0-0.1     &   324   &  $+$0.002(0.967)       &   $-$0.041(0.462)     &  $-$0.009(0.873)          &  $-$0.037(0.504)     &  $-$0.182(0.026)            \\  
     All         &                   &   619    &  $+$0.015(0.696)      &   $+$0.052(0.182)     &  $-$0.083(0.039)         &  $+$0.065(0.095)     &  $-$0.119(0.009)         \\  \hline 
    NLS1     &                  &   1031    &  $-$0.042(0.168)       &   $-$0.006(0.839)     &  $-$0.045(0.146)        &  $+$0.052(0.079)     & $+$0.017(0.572)        \\
    BLS1     &    0.1-0.2   &    1110   &  $+$0.001(0.972)       &   $-$0.026(0.375)     &  $+$0.033(0.283)        &  $-$0.028(0.345)     &  $-$0.006(0.881)        \\  
     All         &                  &     2141  &  $-$0.019(0.356)      &   $+$0.074(4e-04)     &  $-$0.120(3e-08)         &  $+$0.068(1e-03)     &  $-$0.040(0.099)        \\  \hline 
   NLS1     &                   &   1453    &  $-$0.037(0.138)       &   $-$0.006(0.785)     &  $-$0.021(0.412)        &  $+$0.001(0.979)      & $-$0.038(0.137)         \\
    BLS1     &   0.2-0.3    &    1539   &  $-$0.012(0.638)       &   $+$0.020(0.429)    &  $-$0.047(0.068)        &  $+$0.010(0.687)      &  $+$0.058(0.029)            \\  
     All         &                  &    2992   &  $-$0.025(0.164)       &   $+$0.102(1e-08)    &  $-$0.139(4e-14)        &  $+$0.045(0.012)     &  $-$0.038(0.057)      \\  
    \hline \hline       
    \end{tabular} }
     	\label{Table:corr2}
\end{table*}
\end{center}

\begin{center}    
 \begin{table*}     
 \caption{Correlation of amplitude of variability ($\sigma_d$) in z band with different parameters as Table 1. The values given in columns 4-8 are the Spearman correlation coefficient (the $p$-value of no correlation).}
 	\resizebox{\linewidth}{!}{%
     \begin{tabular}{ l l r r r r r r r}\hline \hline   
       Sample & redshift    & Size  &  $\log \lambda L_{5100}$   &   $\log M_{\mathrm{BH}}/M_{\odot}$   &  $\log \lambda_{\mathrm{Edd}}$   &  $R_{5007}$   &  $\rm R_{Fe \,II}$     \\   \hline       
    NLS1     &                   & 317      &  $-$0.010(0.857)      &   $-$0.010(0.858)     &  $+$0.004(0.937)        &  $-$0.013(0.816)     & $+$0.031(0.581)       \\
    BLS1     &   0.0-0.1     &  323     &  $+$0.033(0.547)     &   $-$0.042(0.444)     &  $+$0.068(0.218)        &  $-$0.085(0.129)     &  $-$0.098(0.231)        \\  
     All         &                   &  640     &  $+$0.023(0.549)     &   $+$0.054(0.171)     &  $-$0.089(0.023)        &  $+$0.019(0.638)    &  $-$0.077(0.096)        \\  \hline 
    NLS1     &                  &  1085    &  $-$0.023(0.447)      &   $-$0.044(0.141)     &  $+$0.004(0.229)        &  $+$0.010(0.726)     & $+$0.066(0.029)         \\
    BLS1     &    0.1-0.2   &  1081    &  $-$0.011(0.708)      &   $+$0.008(0.775)     &  $+$0.014(0.624)       &  $-$0.027(0.371)     &  $+$0.053(0.210)            \\  
     All         &                   &  2166   &  $+$0.001(0.993)      &   $+$0.093(1e-05)    &  $-$0.125(4e-09)        &  $+$0.056(0.009)     &  $+$0.007(0.750)        \\  \hline 
   NLS1     &                   &  1538     &  $-$0.024(0.335)      &   $-$0.014(0.565)     &  $-$0.022(0.393)        &  $+$0.017(0.499)     & $-$0.025(0.331)        \\
    BLS1     &   0.2-0.3    &  1514     &  $-$0.044(0.081)      &   $-$0.024(0.344)     &  $-$0.013(0.609)        &  $+$0.017(0.497)     &  $+$0.103(0.001)       \\  
     All         &                  &  3052     &  $-$0.034(0.057)      &   $+$0.089(8e-07)    &  $-$0.137(1e-14)        &  $+$0.067(1e-04)     &  $-$0.004(0.839)       \\  
    \hline \hline       
    \end{tabular} }
     	\label{Table:corr2}
\end{table*}
\end{center}

\begin{center}    
 \begin{table*}     
 \caption{Correlation of amplitude of variability ($\sigma_d$) in y band with different parameters as Table 1. The values given in columns 4-8 are the Spearman correlation coefficient (the $p$-value of no correlation).}
 	\resizebox{\linewidth}{!}{%
     \begin{tabular}{ l l r r r r r r r}\hline \hline   
       Sample & redshift    & Size  &  $\log \lambda L_{5100}$   &   $\log M_{\mathrm{BH}}/M_{\odot}$   &  $\log \lambda_{\mathrm{Edd}}$   &  $R_{5007}$   &  $\rm R_{Fe \,II}$     \\   \hline       
    NLS1     &                   &   302    &  $+$0.019(0.739)      &   $-$0.020(0.722)     &  $-$0.021(0.722)        &  $-$0.040(0.487)     & $+$0.025(0.668)      \\
    BLS1     &   0.0-0.1     &   314    &  $-$0.016(0.778)       &   $-$0.038(0.502)     &  $-$0.037(0.502)        &  $-$0.098(0.083)     &  $-$0.069(0.411)          \\  
     All         &                   &   616    &  $+$0.006(0.881)      &   $+$0.032(0.426)     &  $+$0.032(0.426)       &  $-$0.019(0.641)    &  $-$0.068(0.418)       \\  \hline 
    NLS1     &                  &   1067    &  $-$0.033(0.272)      &   $-$0.035(0.251)     &  $-$0.035(0.251)        &  $+$0.050(0.099)    & $-$0.019(0.522)       \\
    BLS1     &    0.1-0.2   &   1076    &  $-$0.001(0.962)      &   $+$0.007(0.817)     &  $+$0.007(0.817)       &  $-$0.067(0.025)     &  $-$0.055(0.199)         \\  
     All         &                   &  2143    &  $-$0.017(0.419)      &   $+$0.065(0.002)     &  $+$0.065(0.002)       &  $+$0.035(0.096)     &  $-$0.070(0.005)       \\  \hline 
   NLS1     &                   &  1500     &  $-$-0.010(0.687)     &   $-$0.001(0.984)     &  $+$0.001(0.984)       &  $-$0.001(0.982)      & $-$0.023(0.373)       \\
    BLS1     &   0.2-0.3    &   1506    &  $-$0.038(0.135)      &   $-$0.004(0.888)     &  $-$0.003(0.888)        &  $+$0.013(0.615)     &  $-$0.071(0.025)         \\  
     All         &                  &   3006    &  $-$0.026(0.141)      &   $+$0.089(8e-07)    &  $+$0.089(8e-07)       &  $+$0.047(0.009)     &  $-$0.012(0.521)        \\  
    \hline \hline       
    \end{tabular} }
     	\label{Table:corr2}
\end{table*}
\end{center}


\bsp	
\label{lastpage}
\end{document}